\documentclass[12pt]{article}
\usepackage{amsmath,amssymb,amsthm,amsfonts,graphicx}
\numberwithin{equation}{section}
\numberwithin{figure}{section}
\date{\today}
\newcommand{\nc}{\newcommand}
\nc{\nit}{\noindent}
\nc{\D}{\partial}
\nc{\diff}[2]{\frac{d #1}{d #2}}
\nc{\diffn}[3]{\frac{d^{#3} #1}{d {#2}^{#3}}} 
\nc{\pdiff}[2]{\frac{\partial #1}{\partial #2}} 
\nc{\pdiffn}[3]{\frac{\partial^{#3} #1}{\partial{#2}^{#3}}} 
\nc{\abs}[1] {\lvert #1 \rvert} 
\nc{\cE}{{\cal E}} 
\nc{\cZ}{{\cal Z}} 
\nc{\cT}{{\cal T}} 
\nc{\order}{{\cal O}}
\nc{\Eplus}{E_+} 
\nc{\Eminus}{E_-} 
\nc{\Epm}{E_\pm}
\nc{\eplus}{e_+} 
\nc{\eminus}{e_-} 
\nc{\epm}{e_\pm}
\nc{\eps}{\epsilon} 
\nc{\veps}{\varepsilon} 
\nc{\G}{\Gamma} 
\nc{\w}{\omega} 
\nc{\tlambda}{\tilde\lambda} 
\renewcommand{\k}{\kappa}
\renewcommand{\d}{\delta}
\renewcommand{\l}{\lambda}
\newtheorem*{hypo}{Hypothesis}
\nc{\g}{\gamma} 
\newcommand{\n}{\nu}

\nc{\pZ}{\partial_Z} 
\nc{\pT}{\partial_T} 
\nc{\pz}{\partial_z}
\nc{\pt}{\partial_t} 
\nc{\infint}{\int_{-\infty}^{\infty}}
\DeclareMathOperator{\sech}{sech}

\nc{\halfwidth}{6.5cm}
\nc{\figwidth}{10cm}

\begin{document}
\title{Stopping Light on a Defect} 
\author{R.H. Goodman%
\thanks{Mathematical Sciences Research, Bell Laboratories--Lucent
Technologies, Murray Hill, NJ 07974, and Program in Applied and Computational
Mathematics, Princeton University, Princeton, New Jersey} 
\hspace{.05 in}
R.E. Slusher%
\thanks{Optical Physics Research, Bell Laboratories--Lucent Technologies,
Murray Hill, NJ 07974} 
\hspace{.05 in}
M.I. Weinstein%
\thanks{Mathematical Sciences Research, Bell Laboratories--Lucent
Technologies, Murray Hill, NJ 07974}
} 
\maketitle

\begin{abstract}
\textit{Gap solitons} are localized nonlinear coherent states which have been
shown both theoretically and experimentally to propagate in periodic
structures. Although theory allows for their propagation at any speed $v$,
$0\le v\le c$, they have been observed in experiments at speeds of
approximately $50\%$ of $c$. It is of scientific and technological interest to
trap gap solitons. We first introduce an explicit multiparameter family of
periodic structures with localized defects, which support linear defect modes.
These linear defect modes are shown to persist into the nonlinear regime, as
{\it nonlinear defect modes}.  Using mathematical analysis and numerical
simulations we then investigate the capture of an incident gap soliton by
these defects.  The mechanism of capture of a gap soliton is resonant transfer
of its energy to nonlinear defect modes.  We introduce a useful bifurcation
diagram from which information on the parameter regimes of gap soliton
capture, reflection and transmission can be obtained by simple
conservation of energy and resonant energy transfer principles.

\end{abstract}

\section{Introduction}

Solitons are important carriers of energy in many physical systems.  The
emergence of solitons is understood as a consequence of the balance of
dispersive and nonlinear effects on the same length scale.  Optical temporal
solitons~\cite{HT,MSG} and more recently dispersion-managed
solitons~\cite{Agrawal} have been considered candidates for the \textit{bits}
with which to transfer information over long distances. Recent advances in
fabrication of optical fiber with microstructure have rendered the possibility
of storing information in the form of optical \textit{gap solitons} a natural
direction for investigation.

Gap solitons are nonlinear bound states which propagate in periodic
structures.  These have been anticipated in theoretical work~\cite{AW:89,
DS:94}, and observed in experiments~\cite{BRI, CM:87,E:97, M_etal} on
sufficiently high intensity light propagation in optical fiber with a
periodically varying refractive index (a uniform fiber grating). In contrast
to bare fiber used in long distance communications, where the formation length
for solitons is on the order of kilometers, the formation length for gap
solitons is on the order of centimeters.  In theory, gap solitons can travel
with any speed $v$, with $0\le v\le c$, where $c$ denotes the speed of
light. Experiments have demonstrated the slowing of gap solitons to about
$50\%\ c$.

Gap solitons propagate in fibers with uniform grating structures.  In this
paper we examine gratings with localized defects to the amplitude and phase of
the grating. We ask whether it is possible to trap moving gap solitons at the
defect location.  If so, one can envision this having important technological
applications, \textit{e.g.} optical buffers or optical memory.  Through a
careful series of numerical experiments we show how it is possible to trap gap
solitons at a defect and elucidate the mechanism by which light energy is
trapped. A similar question is studied in~\cite{BD:98} using a point-particle
model for the gap soliton/defect interaction.  We compare our results with the
conclusions drawn in that study.  Although we refer to gap soliton capture, it
is perhaps better called capture of gap soliton energy, for it involves the
transfer of the gap soliton's energy to a nonlinear defect mode.

This paper is laid out as follows.  In Section~\ref{sec:cmtheory}, we derive a
variable coefficient version of the nonlinear coupled mode equations from an
appropriate one-dimensional nonlinear Maxwell model. In
Section~\ref{sec:gap_soliton}, we review a few facts about the gap soliton. We
derive a family of defects which support linear bound states (defect modes)
for the coupled mode equations in Section~\ref{sec:linear_defects}.  We then
examine how these bounds states are deformed in the presence of nonlinearity
(nonlinear defect modes) in Section~\ref{sec:nonlinear_defect_modes}. The
results of this study are encoded in bifurcation diagrams that display the
intensity as a function of the frequency for \textbf{(a)} nonlinear defect
modes and \textbf{(b)} gap solitons.  With the aid of these diagrams, we
develop a criterion for trapping and an understanding of its efficiency based
on the notions of resonant energy transfer and energy conservation.  Guided by
this analysis, in Section~\ref{sec:simulations}, we perform a careful series
of numerical experiments to show how the nonlinear bound states interact with
the gap soliton to trap light energy. Simulations are carried out for the
nondimensional system~\eqref{eq:nlcme_nondim}.  Dimensional experimental
parameters are displayed in Appendix~\ref{sec:gscalc}.
Section~\ref{sec:nldamping} contains a brief discussion of the effect of
nonlinear damping, a non-negligible effect in certain highly nonlinear
materials, on soliton propagation and trapping.  A summary and discussion of
results are given in Section~\ref{sec:summary}.  In Appendix~\ref{sec:gscalc}
we give physical parameters for silica fiber~\cite{E:97}, discuss
nondimensionalization and tabulate the dimensional values of parameters
corresponding to the simulations described in
Section~\ref{sec:simulations}. In Appendix~\ref{sec:general_defect}, we
describe a method for deriving defects supporting linear bound states with
prescribed characteristics.

\section{Coupled Mode Theory in a Grating with Defects}
\label{sec:cmtheory}
We consider propagation of light in one dimension in an optical fiber with a
refractive index which is a spatially localized perturbation about a uniformly
periodic index. We model the propagation of low intensity light, confined to a
core mode of the fiber by the wave equation:
\begin{equation}
\D_t^2\left[\ n^2(z)E(z,t)\ \right]\ =\ c^2\ \D_z^2\ E, \label{eq:waveequation}
\end{equation}
where the refractive index is given by:
\begin{equation}
n = \Bar n + 
\Delta n\bigl( \frac{1}{2}W(z)\ +\ \n(z) \cos{(2k_B z + 2 \Phi(z))} \bigr).
\label{eq:defect_def}
\end{equation}
Here, $\Bar n$ denotes the refractive index of the bare fiber and $\Delta n$,
the index 
contrast, is assumed small.  The functions $\n$, $\Phi$, and $W$ model the
defect and are assumed to vary slowly compared to the rapid sinusoidal
variation of the refractive index.  A spatially localized deviation from a
uniformly periodic structure of period
\begin{equation}
d=\pi/k_B
\label{eq:bragg}
\end{equation}
 is obtained by taking 
\begin{align}
\n(z)&\to 1, \label{eq:nulimit}\\ 
\D_z\Phi(z)&\to 0,\ {\rm and}\ 
\D_zW(z)\to 0,\ {\rm as}\  \abs{z}\to\infty.
\end{align}
\begin{figure}
\begin{center}
\includegraphics[width=\figwidth]{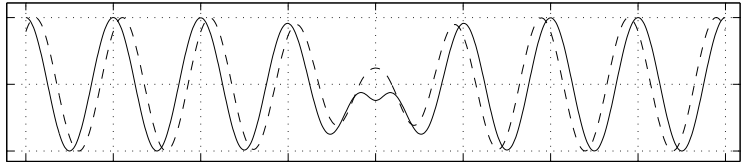}
\caption{Solid and dashed curves are two different periodic index profiles 
with localized defects having the same ``spectral characteristics'' (see
Section~\ref{sec:DSdefect}).} 
\label{fig:schematic}
\end{center}
\end{figure}
Low intensity light propagating in the bare fiber ($\Delta n\ =\ 0$) is
governed by the spatially homogeneous linear wave equation which supports
independently propagating, forward and backward plane wave solutions $E_\pm
e^{i(kz-\w t)}$, where
\begin{equation} 
\w = \pm\frac{ck}{\Bar n}.
 \label{eq:dispersion_relation}
\end{equation} 

The periodic structure ($\Delta n\ne0$) couples these backward and forward
components. This effect is most pronounced for wavelengths in the medium at or
near $\l=2d$ or equivalently the (free space) \textit{Bragg wavelength}
\begin{equation} \lambda_B = 2
\Bar n d. \label{eq:braggwavelength} 
\end{equation} 

For modeling propagation in the nonlinear regime we assume an instantaneous
nonlinear polarization~\cite{A:95}:
\begin{equation}
P_{\rm NL}= \eps_0 \chi^{(3)} E^3.\label{eq:PNL}
\end{equation}
Combining this with~\eqref{eq:defect_def}, the squared index of refraction
with linear and nonlinear effects included is then
\begin{equation}
n^2(z,E^2)\ =\ \Bar{n}^2 + \Bar n \Delta n W(z)
+ 2\Bar n \Delta n\n(z) \cos{(2 k_B z + 2 \Phi(z))}
+ \chi^{(3)} E^2.
\label{eq:nonlinear_index}
\end{equation}
For high intensities the electric field evolves under the nonlinear wave equation%
\footnote{We work with the
model~\eqref{eq:nonlinear_index}--\eqref{eq:maxwell_dim} since it yields a
simple derivation of the envelope equations~\eqref{eq:nlcme1}.  The situation
is however a bit more complicated.
Although~\eqref{eq:nonlinear_index}--\eqref{eq:maxwell_dim} incorporates the
effects of photonic band dispersion, this alone is insufficient to arrest
optical carrier shock formation on the relevant temporal and spatial
scales~\cite{GWH:01}. In fact, a valid envelope description in the absence of
material dispersion would require the incorporation of coupling to
\textit{all} higher harmonics since they are in resonance.  }:
\begin{equation} 
\pt^2\left[ n^2(z,E^2)\ E \right]
= c^2 \pz^2 E.
\label{eq:maxwell_dim}
\end{equation}

So that we can systematically obtain equations of evolution for the forward
and backward carrier wave envelopes, we make explicit our assumptions on the
medium.  We assume that the variation of the refractive index is weak, and
that the deviation from periodicity is small, \textit{i.e.}\ there exists a
small parameter $\veps \ll 1$ such that:
\begin{align*}
\Delta n &= \order(\veps), \\
\pz W = \order(\veps), \pz \nu = \order(\veps), 
\pz \Phi &= \order(\veps),\ {\rm and }\ \pz^2\Phi = \order(\veps^2).
\end{align*}
Due to the periodic structure, we expect coupling of forward and backward wave
components.  This coupling is strongest if the wavelength and period are
chosen according to the above Bragg condition. We now
make a \textit{multiple scales ansatz}, choosing the carrier wavenumber in
\textit{Bragg resonance} with the medium:
\begin{equation}
E = \eplus(z,t) e ^{i(k_B z + \Phi - \w_B t)} 
+ \eminus(z,t) e^{-i(k_B z + \Phi + \w_B t)} + E_1,
\label{eq:ansatz}
\end{equation}
where the wavenumber $k_B$ and frequency $\w_B$ satisfy the dispersion relation
$$
\w_B = \frac{ck_B}{{\Bar n}};
$$
see for example~\cite{GWH:01}.
The first two terms in~\eqref{eq:ansatz} consist of slowly modulated forward
and backward waves. The regime we consider is specified by the above
assumptions on the medium and assumptions on the field amplitude, which we
take to satisfy:
$$
\chi^{(3)}|E|^2\ =\  \order(\veps).
$$
The latter ensures a balance of nonlinearity and \textit{photonic band
dispersion} due to the periodic structure.  We therefore anticipate that the
amplitudes $e_\pm$ will be slowly varying and will satisfy:
\begin{align*}
\pt \epm &= \order(\veps),\ \pz \epm = \order(\veps), \\
\pt^2 \epm &= \order(\veps^2),\ \pz^2 \epm = \order(\veps^2),
\end{align*}
The envelope functions $e_\pm$ in~\eqref{eq:ansatz} are finally determined by
the constraint that the correction terms are of higher order in $\veps$ over a
time scale and length scale of order $\order(\veps^{-1})$,
\begin{equation} 
E_1/\epm = \order(\veps).
\label{eq:smallerror}\end{equation}
The condition~\eqref{eq:smallerror} requires the removal of resonant forcing
terms in the equation for $E_1$. This is equivalent to the constraint that
$\epm$ satisfy the variable coefficient nonlinear coupled mode equations:
\begin{equation}
\begin{split}
i \frac{\Bar n}{c}\D_t\eplus + i \D_z\eplus + \Tilde V(z)\eplus 
+ \Tilde \k(z) \eminus + \Tilde \G(\abs{\eplus}^2 + 2\abs{\eminus}^2)\eplus
&=0 \\
i \frac{\Bar n}{c}\D_t\eminus - 
 i \D_z\eminus + \Tilde V(z)\eminus 
+ \Tilde \k(z) \eplus + \Tilde \G(\abs{\eminus}^2 + 2\abs{\eplus}^2)\eminus
&=0. 
\label{eq:nlcme1}
\end{split}
\end{equation}
The coefficient functions in~\eqref{eq:nlcme1} are defined in terms of the 
parametric functions which characterize the index
profile~\eqref{eq:defect_def}.%
\footnote{It is common to slightly redefine $\tilde \k$ by 
$$
\tilde\k(z) = \eta\cdot\pi \Delta n \nu(z)/\lambda_B
$$
where $0<\eta<1$ is defined as an overlap integral of the radial variation of
the forward and backward modes and represents the fraction of total energy in
the core of the fiber.}
\begin{align}
\Tilde \k(z) &= \frac{\pi \Delta n}{\lambda_B} \n(z)\label{eq:K_def}\\
\Tilde V(z) &= \frac{\pi \Delta n}{\lambda_B}  W(z) - \Phi'(z)
\label{eq:V_def}\\ 
\Tilde \G &= \frac{3 \pi \chi^{(3)}}{\Bar n \lambda_B}. \label{eq:G_def}
\end{align}
Our point of view is to specify grating parametric functions $W(z),\ \nu(z)$
and $\Phi(z)$ through the constitutive law~\eqref{eq:defect_def}. These
determine the functions $\Tilde\k(z)$ and $\Tilde V(z)$, arising in the
coupled mode equations, governing the nonlinear propagation. Note the
appearance of the combination of $W$ and $\Phi'$ in~\eqref{eq:V_def}.
Therefore, spectral characteristics arising due to a DC variation ($W$) in the
index can be, within this approximation, equivalently achieved through phase
variations ($\Phi$).  The solid and dashed curves in
Figure~\ref{fig:schematic} are of index profiles which are equivalent in this
sense; see also Section~\ref{sec:DSdefect}.

The assumptions on the variable coefficients guarantee that away from the
defect, the system approaches the constant coefficient NLCME%
\begin{equation}
\Tilde V(z) \to 0 \ {\rm  and  }\  
\Tilde \k(z) \to  {\Tilde \k}_\infty \ \equiv\   \frac{\pi\Delta n}{\lambda_B}.
\label{eq:t_kappa_infinity}
\end{equation}
We introduce typical dimensional length, $\cZ$, time, $\cT= \cZ\Bar n/c$, and
electric field, $\cE$.  Using these, we define \textit{nondimensional}
spatial and temporal variables $Z$ and $T$, and electric field $\Epm$ given
by:
\begin{equation}
z= \cZ Z, \quad t = \frac{\cZ \Bar n}{c} T, \quad \ {\rm and}\ \quad
\epm = \cE \Epm.
\end{equation}
Then~\eqref{eq:nlcme1} can be expressed in nondimensional form as
\begin{equation} \label{eq:nlcme_nondim} 
\begin{split}
i\D_T\Eplus + i \D_Z\Eplus  
+ \k(Z) \Eminus + V(Z)\Eplus
+ \G(\abs{\Eplus}^2 + 2\abs{\Eminus}^2)\Eplus
&=0 \\
i\D_T\Eminus - i \D_Z\Eminus 
+ \k(Z) \Eplus + V(Z) \Eminus
+ \G(\abs{\Eminus}^2 + 2\abs{\Eplus}^2)\Eminus
&=0,
\end{split}
\end{equation}
where
\begin{equation}
\k(Z) = \cZ \Tilde \k(\cZ Z), \;
V(Z) = \cZ \Tilde V(\cZ Z),
\text{ and }
\G = \cZ \cE^2 \Tilde \G. 
\label{eq:nd_def}
\end{equation}

Our analysis and computer simulations are carried out for the nondimensional
system~\eqref{eq:nlcme_nondim}.  Conversions to dimensional form are given for
important quantities in Appendix~\ref{sec:gscalc}.  Note that $\Tilde\G$ and
therefore $\G$ are positive; see~\eqref{eq:G_def}.

\section{The Gap Soliton}
\label{sec:gap_soliton}
The linearized constant coefficient NLCME ($\k=\k_\infty,\ V=0$) have a
``gap'' in their spectrum. It has no plane-wave solutions with frequencies in
the range $(-\k_\infty,\k_\infty)$; see Section~\ref{sec:rad_modes}. The
nonlinear equations support a family of traveling pulses called a gap
solitons~\cite{AW:89,CJ:89}. The family is parameterized by a velocity $v$ and
a detuning parameter $\d$ with $\abs{v}<1$ and $0 \le \d \le \pi$.  It is
given by:
\begin{equation}
\label{eq:gapsoliton}
\begin{split}
\Eplus &=  s\alpha e^{i\eta} 
\sqrt{\left|\frac{\k_\infty}{2\G}\right|}
\frac{1}{\Delta}
\left(\sin{\d}\right) e^{i s\sigma} \sech{(\theta - i \d/2)}\enspace;\\
\Eminus &= - \alpha e^{i\eta} 
\sqrt{\left|\frac{\k_\infty}{2\G}\right|}
\Delta   
\left(\sin{\d}\right) e^{i s\sigma} \sech{(\theta + i \d/2)}\enspace;
\end{split}
\end{equation}
where:
\begin{gather*}
\g = \frac{1}{\sqrt{1-v^2}} \enspace ; \qquad
\Delta = \left(\frac{1-v}{1+v}\right)^{\frac{1}{4}}\enspace ; \\
\theta = \g \k_\infty (\sin{\d})(z - v t)\enspace ; 
\qquad
\sigma = \g \k_\infty (\cos{\d})(v z - t)\enspace ; 
\\
\alpha = \sqrt{\frac{2(1-v^2)}{3-v^2}}\enspace ;  \qquad
s = {\rm sign}(\k_\infty \G)
\\
e^{i\eta}= \left( - \frac
{e^{2\theta} + e^{- i \d}}
{e^{2\theta} + e^{i \d}} \right)^{\frac{2v}{3-v^2}}.
\end{gather*}

\begin{figure}
\begin{center}
\includegraphics[width=\halfwidth]{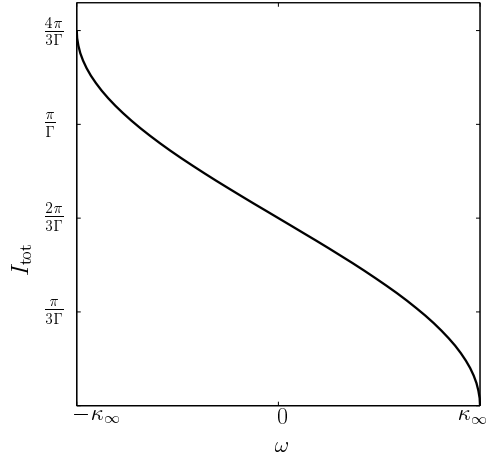}
\caption{The intensity of the stationary gap soliton as a function of its
frequency.} 
\label{fig:gs_freq_L2}
\end{center}
\end{figure}

The temporal frequency of the gap soliton is $\k_\infty\g\cos\d$ which is
inside the band gap for $\abs{v}<\sin\delta$, although in a reference frame
moving at speed $v$, the frequency is always in the band gap.  We define the
maximum intensity $I_{\rm max}$ and the total intensity $I_{\rm tot}$ and give
their values for the gap soliton:
\begin{align}
I_{\rm max} &= \max_Z{(\abs{\Eplus}^2+\abs{\Eminus}^2)}
 = \frac{8\k_\infty \sqrt{1-v^2}}{\Gamma(3-v^2)} 
\sin^2\frac{\d}{2} , \label{eq:Imax}\\
I_{\rm tot} &= \infint(\abs{\Eplus}^2+\abs{\Eminus}^2)\,dZ
 =\frac{4(1-v^2)\d}{\Gamma(3-v^2)}, \nonumber 
\end{align}
and the full width at half-maximum (FWHM) by
\begin{equation}
{\rm FWHM} = \frac{2\sqrt{1-v^2}}{\k_\infty \sin \d}
\cosh^{-1}\sqrt{1+\cos^2{\frac{\d}{2}}}. 
\label{eq:FWHM}
\end{equation}
In mathematical analysis, the square root of the total intensity $I_{\rm tot}$
is often referred to as the $L^2$ norm.  In Figure~\ref{fig:gs_freq_L2}, we
plot the intensity as a function of frequency for the stationary ($v=0$) gap
soliton.  This curve is parameterized by
$$
\w(\d)=\k_\infty \cos \d, \; I_{\rm tot}(\d)= \frac{4\d}{3\Gamma} \quad
\text{for}\, \d \in [0,\pi]. 
$$
Results on linearized stability and instability of gap
solitons in different parameter regimes are obtained in~\cite{BPZ}.

\section{Defects and Linear Defect Modes}
\label{sec:linear_defects}

The functions $V(Z)$ and $\k(Z)$ define a defect in our periodic medium
and we now seek the linear modes associated with this defect. These are
solutions of the linear coupled mode equations:
\begin{subequations} \label{eq:lcme}
\begin{align}
i\D_T\Eplus + i \D_Z\Eplus
+ \k(Z) \Eminus + V(Z)\Eplus
&=0 \\
i\D_T\Eminus - i \D_Z\Eminus
+ \k(Z) \Eplus + V(Z) \Eminus
&=0,
\end{align}
\end{subequations}
or
\begin{equation}
\left[\ i\D_T\ +\  i\sigma_3\D_Z\ +\ V(Z)\ +\ \k(Z)\sigma_1\ \right]E=0,
\label{eq:lcme1}
\end{equation}
where 
\begin{equation}
E\ =\ \begin{pmatrix} E_+\\ E_-\end{pmatrix},\
  \sigma_1\ =\ \begin{pmatrix} 0 & 1 \\ 1 & 0\end{pmatrix},
 \ \sigma_3\ =\ \begin{pmatrix} 1 & 0\\ 0 & -1\end{pmatrix}.
\label{eq:Esigma13}
\end{equation}

Substitution of the Ansatz
\begin{equation} 
E(Z) =\ e^{-i\w T}\  e^{i\sigma_3\int_0^ZV(\zeta)d\zeta}\ F(Z)
\label{eq:Eansatz}
\end{equation}
yields
\begin{align}
\D_Z F &= \begin{pmatrix} i\w & u(Z) \\ 
                 \overline{u(Z)} & -i\w \end{pmatrix}\ F,
          \label{eq:ZS}\\
u(Z) &= i\k(Z)e^{-2i\int_0^Z V(\zeta)\ d\zeta}\label{eq:udef}
\end{align} 
Solutions of the form~\eqref{eq:Eansatz}, which are square integrable in $Z$
are called \textit{defect modes}. Solutions of the form~\eqref{eq:Eansatz}
which are bounded and oscillatory in $Z$ are called \textit{radiation modes}. 
The set of frequencies, $\w$, corresponding to defect modes and radiation
modes is called the \textit{spectrum} of~\eqref{eq:ZS}.

One can pose the question: can prescribed spectral characteristics
of~\eqref{eq:lcme} (\textit{e.g.}\ defect modes and reflection and
transmission spectra) be achieved by appropriate choice of $u(Z)$ ($\k(Z)$ and
$V(Z)$)? This question was considered in~\cite{SS:85} to study grating and
filter design using the Gelfand-Levitan-Marchenko approach to inverse
scattering; see also~\cite{W:99}. The Gelfand-Levitan-Marchenko method, as
used in~\cite{ZS1,SS:85}, can be used to characterize gratings with desired
spectral characteristics.

\subsection{Radiation modes for a general localized defect}
\label{sec:rad_modes}
We suppose that the function $u$ in~\eqref{eq:ZS} has the asymptotic behavior:
$$
 u(Z)\to\rho e^{i\theta_\pm},\ Z\to\pm\infty
$$
corresponding to a spatially localized defect in the periodic structure.

The values of $\w$ lying in the continuous spectrum are then characterized
by the equation:
$$
\D_Z  F \ =\  \begin{pmatrix}i\w & \rho e^{i\theta_\pm}\\
                \rho e^{-i\theta_\pm} & -i\w\end{pmatrix}\  F.
$$

Seeking solutions of the form: $e^{iQZ}\vec v$, with $Q$ real, we find that
there are nontrivial solutions $\vec v$ provided:
\begin{equation}
\w^2\ =\ \rho^2\ +\ Q^2
\label{eq:disp}
\end{equation}

\begin{figure}
\begin{center}
\includegraphics[width=\halfwidth]{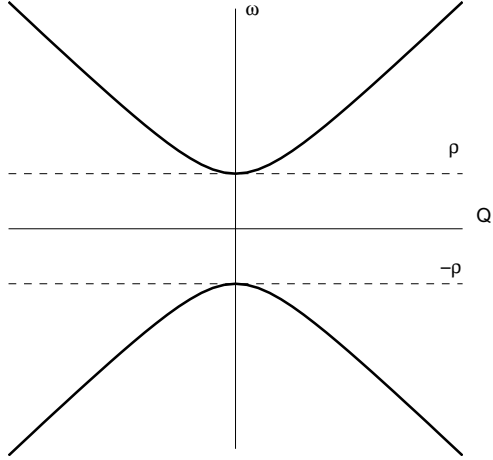}
\caption{The dispersion relation~\eqref{eq:disp}, showing a band gap.}
\label{fig:disp_rel}
\end{center}
\end{figure}

\noindent
Therefore, the continuous spectrum consists of the real axis minus
a gap (photonic band gap): $-\rho < \w < +\rho$; see Figure~\ref{fig:disp_rel}.

\subsection{Dark soliton defect gratings}
\label{sec:DSdefect}
It is interesting to note that the system~\eqref{eq:ZS}, characterizing the
modes of the modulated periodic structure, is the Zakharov-Shabat eigenvalue
problem associated, via the inverse scattering transform, with the
\textit{defocusing} nonlinear Schr\"odinger equation~\cite{ZS1,ZS2}:
\begin{equation}
i\D_\tau u\ -\ \D_Z^2u\ +\ 2|u|^2u.\label{eq:defocusingnls}
\end{equation}
If $u(Z,\tau)$ satisfies~\eqref{eq:defocusingnls} then for each $\tau$, (a)
the spectrum of~\eqref{eq:ZS} is independent of $\tau$ and (b) as $\tau$
varies, the eigenfunctions and radiation modes of~\eqref{eq:ZS} evolve in a
trivial (linear) and explicitly computable manner. That is,
equation~\eqref{eq:defocusingnls} defines an isospectral deformation of the
eigenvalue problem~\eqref{eq:ZS} and therefore, by~\eqref{eq:udef}, provides a
rich class of potentials, $u(Z)$, and therefore modulated gratings
($\k(Z), V(Z)$) with the same spectral characteristics.
 
From the inverse scattering theory of~\eqref{eq:defocusingnls}, we learn that
the bound states and continuum radiation modes associated with~\eqref{eq:ZS}
can be ``mapped'', respectively, to the dark solitons and radiation modes
of~\eqref{eq:defocusingnls}. In this section we explore those grating profiles
which correspond to the simplest dark soliton solutions
of~\eqref{eq:defocusingnls}.

Since the eigenvalue problem for the pair $( F(Z),\w)$ is self-adjoint, $\w$
varies over the real numbers. Furthermore, the set of all $\w$ satisfying
$|\w|\ge\rho$ is \textit{continuous spectrum}. There are no eigenvalues
embedded in the continuous spectrum.  Therefore, if the eigenvalue
problem~\eqref{eq:ZS} has eigenvalues they must occur in the gap $|\w| <
\rho$. Of interest is the following

\noindent{\bf Inverse Problem:} Given $N$ numbers $\w_1,\ldots,\w_N$,
satisfying $|\w_j|<\rho$, find \textit{potentials}, $u(Z)$ for which
these are the eigenvalues of the eigenvalue problem~\eqref{eq:ZS}.

This inverse problem has many solutions.  A remarkable class of solutions are
those, for which the reflection coefficient associated with~\eqref{eq:ZS} is
zero.  These are the \textit{dark N-soliton} solutions of defocusing NLS.

\noindent\textbf{Dark solitons (N=1):}
Let $k\ne0$ be arbitrary and $|\w|<\rho$. Let
\begin{equation*} \rho \ =\ \abs{\w + ik}.
\end{equation*}
Define
\begin{equation}
u(Z) = e^{i\phi}( \w- i k  \tanh{(kZ)}\ )
\label{eq:darksoliton}
\end{equation}
where $\phi$ is left unspecified to this point.

It can be verified easily that $\l$ is an eigenvalue of~\eqref{eq:ZS} with
corresponding eigenfunction:
\begin{equation}
 F(Z)\ =\
\binom{1}{ie^{-i\phi}}\sech(kZ),
\label{eq:defectmode}
\end{equation}
\bigskip
Therefore, $E_\pm$ are given by: 
\begin{equation}
\binom{\Eplus}{\Eminus} =
\binom {e^{\frac{i}{2} \arctan{\frac{k\tanh{(kZ)}}{\w}}}}
       {i e^{-i\phi}e^{-\frac{i}{2} \arctan{\frac{k\tanh{(kZ)}}{\w}}}}
e^{-i\w t}  \sech{(kZ)}.
\label{eq:Epm}
\end{equation}

\noindent\textbf{Specification of the grating with prescribed defect energy:}

A grating is specified by the functions $V(Z)$ and $\k(Z)$
(see~\eqref{eq:V_def} and~\eqref{eq:K_def} and~\eqref{eq:nd_def}).
Using~\eqref{eq:udef} we obtain a relation between a family of ``dark
solitons'', $u(Z)$, and the functions $V$ and $\k$:
\begin{equation}
 i\k(Z) e^{-2i\int_0^Z V(s) ds} =
 e^{i\phi}\left(\w -ik\tanh{(kZ)}\right).
\label{eq:kappaVu_relation}
\end{equation}
Choosing $\phi=\pi/2$, this yields
\begin{subequations}
\label{eq:defectsolve}
\begin{align}
\k(Z) &= \left[\w^2 +
                k^2\tanh^2{(kZ)}
                \right]^\frac{1}{2},\label{eq:kappasolve}\\
 V(Z)  &= \frac{1}{2} k^2\ \w\ \left[ \w^2 +
                 k^2\tanh^2{(kZ)}\right]^{-1} 
            \ \sech^2{(kZ)}   \label{eq:Vsolve}
\end{align}
\end{subequations}
and sets the term
$$
ie^{-i\phi}=1
$$ 
in equations~\eqref{eq:defectmode} and~\eqref{eq:Epm}. Note that the limit
$\w\to 0$ of the defect definition~\eqref{eq:defectsolve} is singular:
\begin{subequations}
\begin{align}
\k(Z) &= \abs{k\tanh(kZ)},\\
V(Z) &= \pm\frac{\pi}{2} \delta(Z)
\end{align}
\label{eq:zerolimit}
\end{subequations}
where $\d(Z)$ denotes the Dirac delta function and the sign on $V(Z)$ depends
on the direction on which the limit $\w\to 0$ is taken.
Instead, taking $V(Z)=0$, $\w=0$, $e^{i\phi} = \pm 1$ in
equation~\eqref{eq:kappaVu_relation} yields the continuous limit
\begin{align}
\k(Z) &= - k \tanh{kZ} \label{eq:kappazero}\\
F(Z) &= \binom{1}{\mp i} \sech{kZ}
\label{eq:Fzero}
\end{align}

Recall that in our coupled mode system, the medium is characterized by three
functions: $\nu$, $W$, and $\Phi$. From our solution we see
by~\eqref{eq:K_def} that~\eqref{eq:kappasolve} (or~\eqref{eq:kappazero})
uniquely determines $\nu$. However, from~\eqref{eq:V_def} we see
that~\eqref{eq:Vsolve} specifies only a linear combination of $W$ and $\Phi'$,
giving one some freedom in how to design a medium with the desired spectral
characteristics. In Figure~\ref{fig:schematic} two gratings with identical
$\nu$, $\k$, and $V$ are displayed. The solid curve corresponds to the choice
$\Phi=0$ (no phase shift in the the refractive index), and the dashed curve
corresponds to the choice $W(Z)=0$ (no modulation to the DC component of the
refractive index).

\subsection{More general defect gratings}
\label{sec:more_genl}
\begin{figure}
\begin{center}
\includegraphics[width=\figwidth]{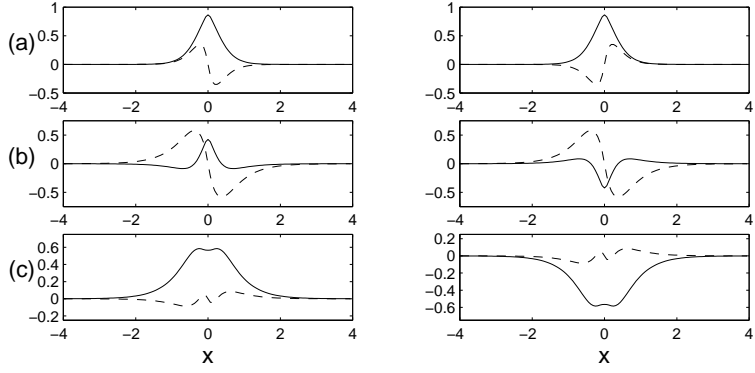}
\caption{The three eigenmodes for the defect~\eqref{eq:gen_defect} with
$(\w,k,n) = (-1,2,2)$.  In each plot $\Eplus$ is in the
left column and $\Eminus$ in the right. Solid and dashed lines correspond to
real and imaginary parts, respectively.} 
\label{fig:eigenmodes}
\end{center}
\end{figure}

In this section we consider a class of defect  gratings which generalizes those
studied in the previous section:
\begin{subequations}
\begin{align}
\k(Z) &= \sqrt{\w^2 + n^2 k^2 \tanh^2 {(kZ)}}; \label{eq:genkappa}\\
V(Z) &=  \frac{\w n k^2 \sech^2{(kZ)}}
               {2(\w^2 + n^2 k^2 \tanh^2{(kZ)})}. \label{eq:genV}
\end{align}
\label{eq:gen_defect}
\end{subequations}
We shall use this class of defects extensively in numerical simulations.  This
family of defects can be obtained by specifying an exponentially localized
eigenfunction of~\eqref{eq:ZS} and then deriving a potential for which this
function is a bound state. The calculation is presented in
Appendix~\ref{sec:general_defect}.  The generalization of the
eigenmode~\eqref{eq:Epm} is
\begin{equation}
\binom{\Eplus}{\Eminus} = 
\binom {e^{\frac{i}{2} \arctan{\frac{n k\tanh{(kZ)}}{\w}}}}
       {e^{-\frac{i}{2} \arctan{\frac{n k\tanh{(kZ)}}{\w}}}}
e^{-i\w t}  \sech^n{(kZ)}. \label{eq:genEpm}
\end{equation}
When $n>1$ in~\eqref{eq:gen_defect}, then the linearized
equations~\eqref{eq:lcme} have multiple bound states.  For example when
$(\w,k,n) = (-1,2,2)$, there are three eigenvalues $\w_0 = -1$ and $\w_{\pm 1}
= \pm \sqrt{13}$.  The eigenfunctions are shown in
Figure~\ref{fig:eigenmodes}. 

The photonic band gap width is determined by the asymptotic behavior of 
$\k$ in~\eqref{eq:genkappa} and is given by $2\k_\infty$, where 
\begin{equation}
\k_\infty\ =\ \lim_{Z\to\infty}\k(Z)\ =\ \sqrt{\w^2+n^2k^2}.
\label{eq:gapwidth}
\end{equation}
In our computer simulations (see Section~\ref{sec:simulations}) we will set
$k^2=k_n^2=C^2/n^2$, thereby fixing the gap width equal to
$2\sqrt{\w^2+C^2}$ and vary the defect width by varying the parameter $n$.
 
The defect $\k(Z)$ varies between the values $\k_0 = \abs{\w}$ and
$\k_\infty= \sqrt{\w^2+ n^2k^2}$.  Therefore the ``depth'' $\Delta_*$
of the defect is given by 
\begin{equation}
\Delta_* = \sqrt{\w^2 + n^2k^2} - \abs{\w}.
\label{eq:depth}
\end{equation}

We define the defect width (FWHM) to be given by twice the value of $Z$ for
which $\k(Z)= \tfrac{1}{2}(\k_0 + \k_\infty)$, yielding
\begin{equation}
{\rm FWHM} = \frac{2}{k} \tanh^{-1} {\left(
\frac{\sqrt{2\abs{\w}\sqrt{\w^2+n^2k^2}+n^2k^2-2\w^2}}{2nk} 
\right)}
\label{eq:defect_fwhm}
\end{equation}
in the nondimensional setting. Dimensional values for the defect depth and
width are provided in Appendix~\ref{sec:gscalc}.

\section{Nonlinear defect modes} \label{sec:nonlinear_defect_modes}

In this section we show that the linear defect modes of
Section~\ref{sec:linear_defects}, upon inclusion of nonlinear terms, deform
into nonlinear defect modes in a sense to made precise below.  We begin by
observing that the dimensionless nonlinear coupled mode
equations~\eqref{eq:nlcme_nondim} can be written in the vector form:
\begin{equation}
\bigl( i\left( \partial_T +  \sigma_3\partial_Z \right) + \sigma_1\k(Z) +  
V(Z) \bigr)E + \G N(E,E^*)E = 0,
\label{eq:vec_nlcme}
\end{equation} 
where $E$ is the two-vector with components $E_\pm$, $\sigma_1$ and $\sigma_3$
are displayed in~\eqref{eq:Esigma13}, and the $N(E,E^*)$ defines the nonlinear
term:
\begin{equation}
N(E,E^*) = 
\begin{pmatrix} 
|E_+|^2+2|E_-|^2 & 0\\ 
0 & |E_-|^2+2|E_+|^2
\end{pmatrix}
\label{eq:Ndef}
\end{equation}

\noindent{\bf Bifurcation of nonlinear defect modes} We assume that all
eigenvalues of the linearized problem~\eqref{eq:lcme} are simple, which has
been found numerically for the family of defects investigated.  Let $E_0 =
e^{-i\w_0T} \cE_0(Z)$ denote a linear defect mode.  That is, $\cE_0(Z)$ is a
spatially localized solution of the equation 
\begin{equation}
\left( \w_0 + i\sigma_3\partial_Z + \sigma_1\k(Z) +
 V(Z) \right)E = 0.
\label{eq:lcme2}
\end{equation} 

We seek to construct \textit{nonlinear} bound states of~\eqref{eq:vec_nlcme}
of the form: 
\begin{equation}
E(Z,T)\ =\ e^{-i \w T}\ \cE(Z),
\label{eq:nl_mode}
\end{equation}
where
\begin{equation}\label{eq:expand}
\begin{split}
\cE(Z) &=  
\alpha\left( E_0(Z)\ + |\alpha|^2 E_1(Z) +  {\cal O}(|\alpha|^4) \right)\\
\w &= \w_0 + \w_1 |\alpha|^2 + {\cal O}(|\alpha|^4)
\end{split}
\end{equation}
and $\alpha$ is a small parameter. Since for any $\w$, $\cE(Z) \equiv 0$ is a
solution of~\eqref{eq:lcme2}, solutions of the form~\eqref{eq:expand} are said
to \textit{bifurcate} from the trivial solution at $\w=\w_0$. 

Substitution of~\eqref{eq:expand} into~\eqref{eq:vec_nlcme} yields a hierarchy
of inhomogeneous linear equations beginning with:
\begin{subequations}
\begin{align}
{\cal O}(1):&\; {\cal L}_0\ E_0\ =\ 0\\
{\cal O}(|\alpha|^2):&\; {\cal L}_0\ E_1\ =\ -\w_1\ E_0\
  -\ \G\ N(E_0,E_0^*)E_0 
\end{align}
\label{eq:hierarchy}
\end{subequations}
where
\begin{equation}
{\cal L}_0\ =\ \w_0\ +\ i\sigma_3\partial_Z\ +\ \sigma_1\k\ +\ V.
\label{eq:Ldef}
\end{equation}
is a linear self-adjoint operator.  The first equation in~\eqref{eq:hierarchy}
holds if $E_0=\cE_0$, any linear defect mode, $\cE_0$ of frequency
$\w_0$. The eigenvalue $\w_0$ is necessarily of multiplicity one.  The
second equation in~\eqref{eq:hierarchy} is solvable for a localized correction
term, $\cE_1$ if and only if the right hand side of the equation is
orthogonal to the null space (zero energy subspace) of ${\cal L}_0$.  Imposing
this orthogonality condition yields the following equation which determines
the value of $\w_1$:
\begin{equation}
\langle \cE_0\ |\ \w_1\ \cE_0\ +\ 
 \G\ N(\cE_0,\cE_0^*) \cE_0
 \rangle = 0. 
\label{eq:orthog}
\end{equation}
We obtain from~\eqref{eq:orthog}:
\begin{equation}
 \w_1 = -\G  \frac{\langle \cE_0 | N(\cE_0,\cE_0^*) \cE_0\rangle}
{\langle \cE_0 | \cE_0 \rangle} 
\label{eq:a_branch}
\end{equation}

It follows that the nonlinear defect mode bifurcating from the linear defect
mode of frequency $\w_0$ is: 
\begin{align}
E(Z,T) &= e^{-i\w T} \alpha\ \left( \cE_0(Z) +\ 
 |\alpha|^2\cE_1(Z) + 
 {\cal O}(|\alpha|^4) 
 \right)\\
\w &= \w_0 + |\alpha|^2 \w_1 + {\cal O}(|\alpha|^4)
\label{eq:Ea_branch}
\end{align}
We shall refer to the nonlinear defect mode bifurcating from $\w=\w_0$ as an
$\w_0$-nonlinear defect mode. Since $\G>0$ (see Section~\ref{sec:cmtheory}),
the bifurcating states have frequencies below $\w_0$.

\begin{figure}
\begin{center}
\includegraphics[width=\figwidth]{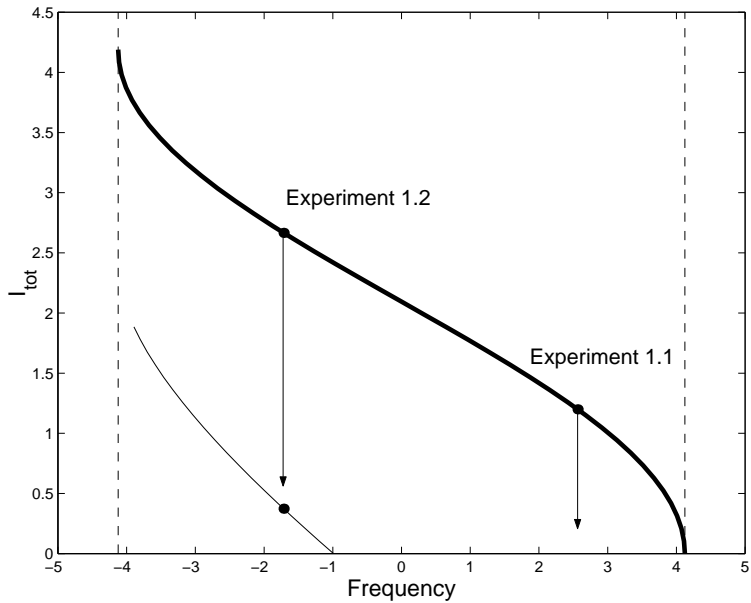}
\caption{Intensity vs.\ frequency for the gap soliton (bold) and a
nonlinear defect mode with parameters $(\w_0,k)=(-1,4)$.}
\label{fig:I_v_freq}
\end{center}
\end{figure}

A rigorous proof of the existence of bifurcating nonlinear defect modes and
the validity of this expansion can be given in a manner analogous to that
carried out in the context of the nonlinear Schr\"odinger
equation~\cite{RW:88}. Numerically we are able to find nonlinear defect modes
for values of $\alpha$ much larger than zero.  A plot of the intensity vs.\
frequency for one such family of defect modes is shown in
Figure~\ref{fig:I_v_freq}.  For the spatially homogeneous case, gap solitons
are seen to bifurcate from the zero state at the right endpoint of the
continuous spectrum, $\w = \k_\infty = \sqrt{17}$. For the given defect, a
branch of nonlinear defect modes bifurcates from the zero state at $\w= \w_0 =
-1$. The nonlinear defect mode and its frequency become difficult to compute
as the frequency $\w$ of the nonlinear mode approaches an endpoint of the
continuous spectrum, because the exponential decay rate decreases and larger
spatial intervals must be used in order to compute the exponential tail.

\section{Computer simulations of a  gap soliton incident on a defect}
\label{sec:simulations}

In Section~\ref{sec:gap_soliton} we discussed gap solitons, the fundamental
nonlinear bound state of propagation in a uniform periodic structure. In
Section~\ref{sec:linear_defects} we then considered the linear modes of a
periodic structure with a defect and, in
Section~\ref{sec:nonlinear_defect_modes}, the nonlinear defect modes which
bifurcate from these linear defect modes. In this section we study by computer
simulation the dynamics of a gap soliton incident on a defect.

In Section~\ref{sec:cmtheory}, we derived a nondimensional form of the coupled
mode equations with nondimensional parameters $\k(Z)$, $V(Z)$, and $\G$ given
by~\eqref{eq:nd_def}.  By an additional rescaling of the nondimensional
variables $Z$, $T$, and $\Epm$, we may redefine the equations so that $\G=1$
and $\k_\infty$ takes the form given in equation~\eqref{eq:gapwidth}.  In
other words, $\k_\infty$ and $\G$ can be set arbitrarily and the results can
always be mapped to a physical system.

Simulations indicate complex interactions between the incident gap soliton
(Section~\ref{sec:gap_soliton}) and the modes of defect
(Sections~\ref{sec:linear_defects} and~\ref{sec:nonlinear_defect_modes}).
\textit{An understanding of the dynamics and the potential for trapping
requires an understanding of the energy exchange between the gap soliton mode
and nonlinear defect modes.}  Our numerical simulations give strong support to
the following hypothesis suggested by the notions of resonant energy transfer
and energy conservation.
\begin{hypo}
Consider a gap soliton incident on a defect with sufficiently low incident
velocity.  The gap soliton will transfer its energy to a nonlinear defect
mode, and thereby be trapped, if there exists a nonlinear defect mode of the
same frequency \textbf{and} lower total intensity ($L^2$ norm).  Otherwise,
the gap soliton energy will be reflected and/or transmitted.
\end{hypo}

We now describe our numerical experiments.  The defects described in
Section~\ref{sec:linear_defects} are members of a 3-parameter ($\w_0$, $k$,
$n$) family and the gap solitons are described by two parameters ($v$,
$\delta$). While it is not possible to investigate all of the soliton-defect
interactions in this 5-dimensional space, we have performed a large number of
simulations, and were able to draw some general conclusions. We concentrate on
gap solitons of comparable width and amplitude to the defects, so that linear
and nonlinear interactions are likely to be strong and balanced.  Physical
experiments have so far produced pulses with relatively large values for $v$
and small values for $\delta$, so we make some attempt to trap gap solitons in
this parameter regime.  Dimensional equivalents for most of the nondimensional
experiments are given in Appendix~\ref{sec:gscalc}.  Intensities range from
130 to 1800 ${\rm GW}/{\rm cm}^{2}$, pulse widths are between 1.3 and 4.4 mm,
and defect widths are between 1.6 and 4.7 mm (FWHM).  Note that in this
section, the frequency of the linear defect mode is given by $\w_0$, while $\w
= \kappa_\infty \cos\d$ is the frequency of a stationary gap soliton.

\subsection{Experiment~1: 
 Gap soliton incident on a dark soliton defect gratings}
\label{sec:dsgrating}

We first consider the simplest case of pulses interacting with the dark
soliton defect gratings defined in~\eqref{eq:defectsolve}.  We consider a
defect with $k=4$ and $\w_0=-1$.  This defect supports a single nonlinear bound state. The
key insight into predicting whether the gap soliton interacts strongly with
the defect is found by examining Figure~\ref{fig:I_v_freq}.  A gap soliton
with frequency $\w = \k_\infty\cos{\d}$ will interact most strongly with
the defect if a nonlinear defect mode exists with the same frequency and equal
or less total intensity.  If it does, then it is possible for the gap soliton
to resonantly transfer its energy to the defect mode. The relevant mechanism
seems not to be the slowing of the soliton, rather this transfer of energy.
Note from the figure that for $\w <-1$ ($\d > 1.82$) such modes exist, and
for larger $\w$ they do not. Of course, the gap soliton frequency-intensity
curve given in the figure applies only in the case $v=0$, but it is useful for
making predictions for small $v$.

\begin{figure}
\begin{center}
\includegraphics[width=\figwidth]{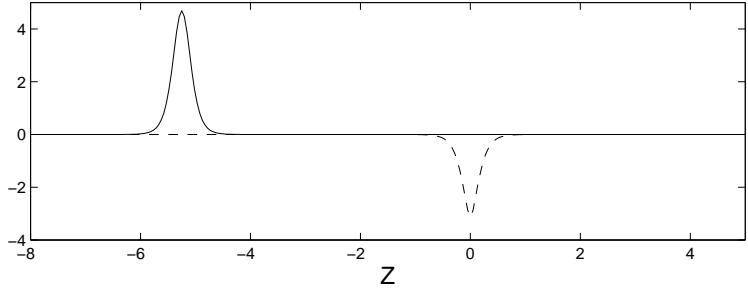}
\caption{(Experiment~1.1) The initial value of $\abs{\Eplus}^2+\abs{\Eminus}^2$
(solid line), which gives the approximate strength of the nonlinear forcing,
and of the defect $\k(Z) - \k_\infty$ (dashed), which gives the
forcing due to the defect.}
\label{fig:initsize}
\end{center}
\end{figure}

\noindent{\bf Experiment~1.1} (Reflection/Transmission)  With the defect
parameters set, we then investigate a two parameter family of gap solitons
indexed by the velocity $v$ and the detuning $\d$.  Our first experiment is
for detuning $\d= 0.9$. Via inspection, the pulse is of comparable depth and
width to the defect, so seems an ideal candidate for capture, see
Figure~\ref{fig:initsize}. 
However, the central frequency of this gap soliton at small velocities is
$\k_\infty \cos \delta \approx 2.56 > \w_0 =-1$, so we do not expect the defect
mode to be strongly forced by the gap soliton. 

Indeed, although we observe a slowing, and therefore delay, of the gap soliton
we do not find significant excitation of the defect mode or trapping.  We find
that below a critical velocity $v \approx 0.257$, all gap solitons are
reflected, and above this speed they are transmitted.  The closer the incoming
pulse comes to this incoming velocity, the longer it remains in the
neighborhood of the defect before being ejected and the velocity of the
outgoing pulse is approximately that of the incoming pulse.

\begin{figure}
\begin{center}
\includegraphics[width=\halfwidth]{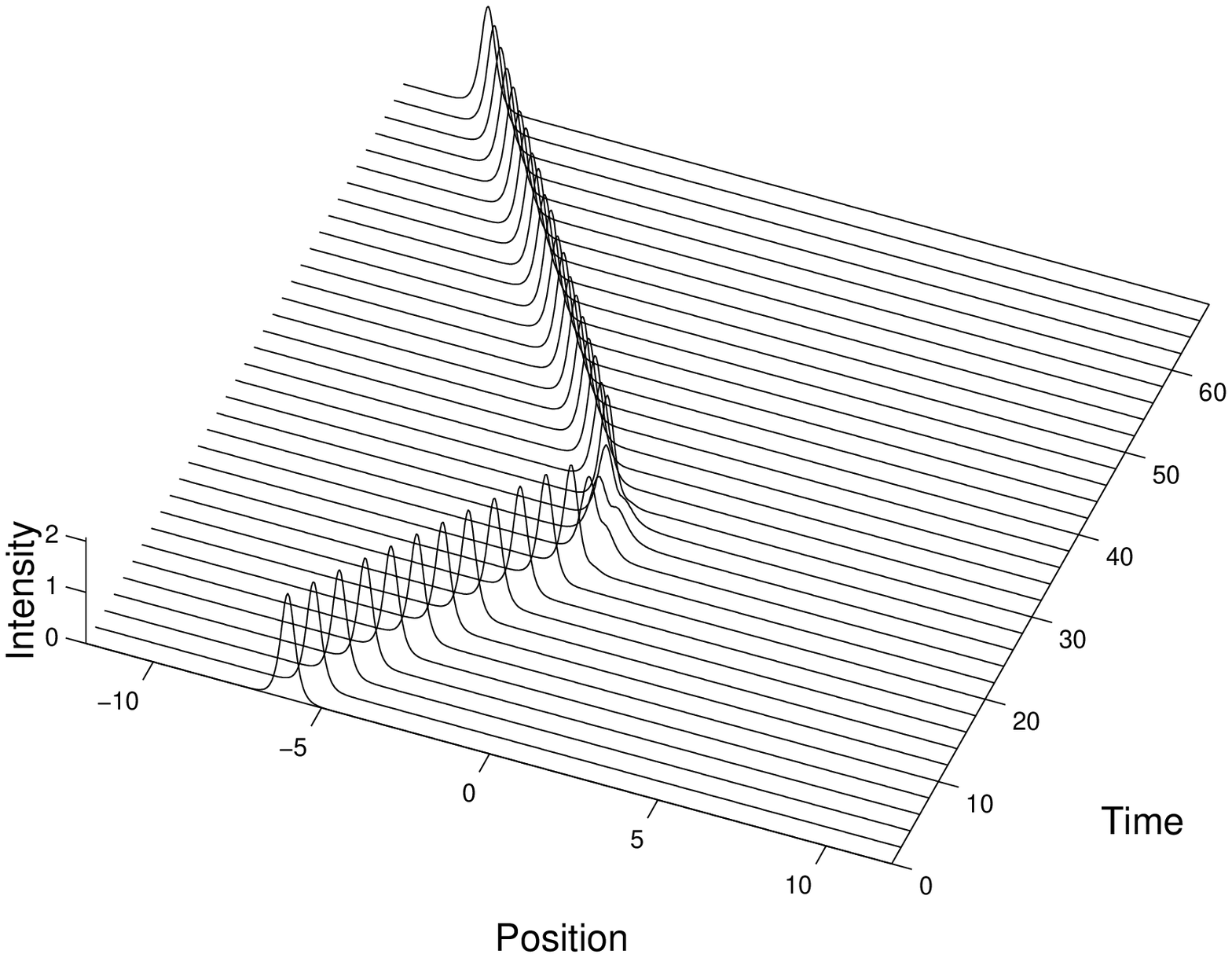}
\includegraphics[width=\halfwidth]{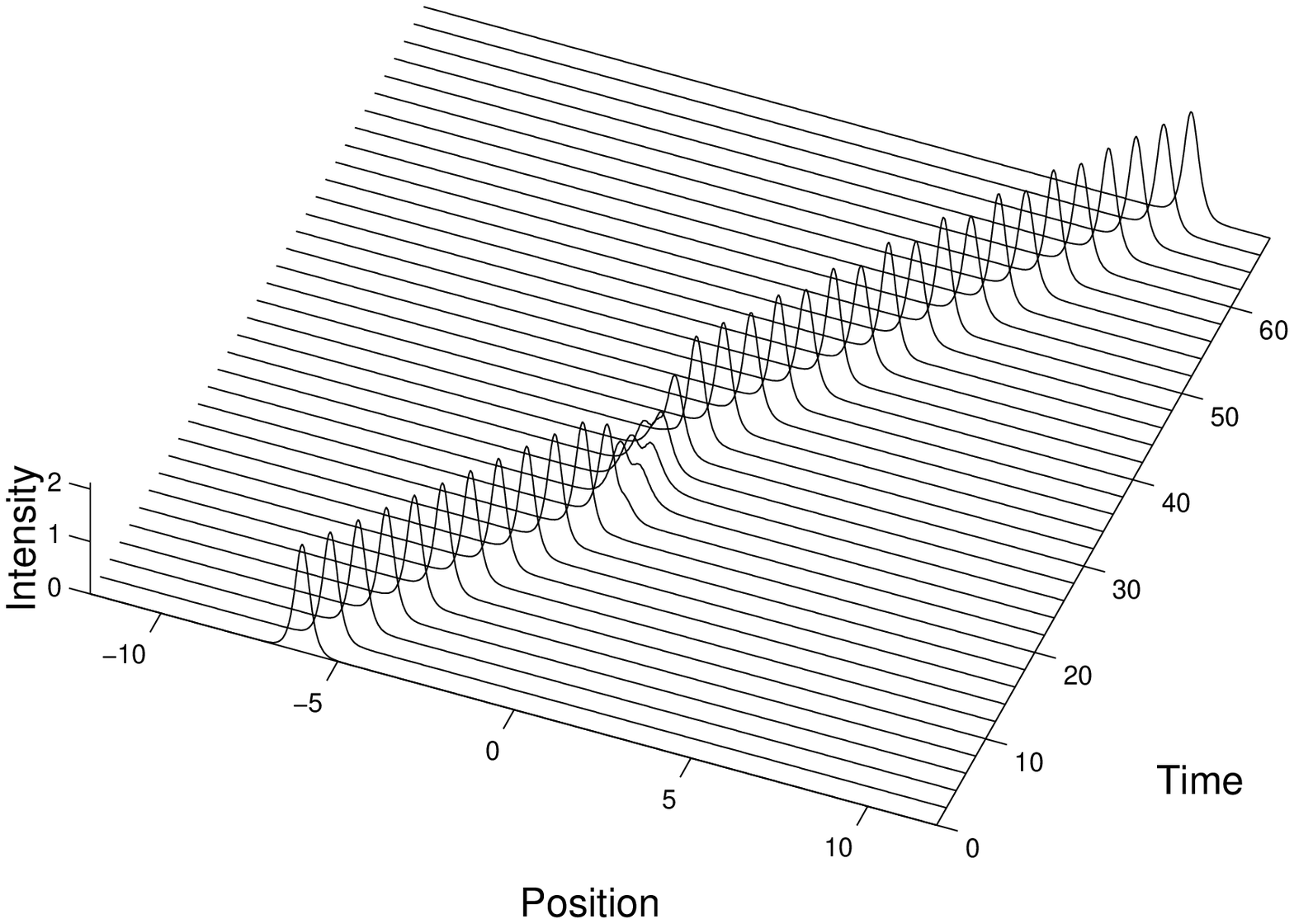}
\caption{(Experiment~1.1) At left, a gap soliton with $v=.2565$ is reflected
by a defect at $Z=0$. At right, a slightly faster gap soliton with $v=.257$ is
transmitted.}
\label{fig:reflected_transmitted}
\end{center}
\end{figure}

In Figure~\ref{fig:reflected_transmitted}, we show the evolution of two gap
solitons incident on the defect, both very close to the critical velocity,
showing clearly the effects discussed above.  

Interestingly, the gap soliton \textit{slows down} when it nears the defect
(Figure~\ref{fig:reflected_transmitted} for times between 20 and 30).
This is somewhat unexpected; as the defect supports a bound state, one
intuitively expects the soliton center of mass to move as a ``classical
particle in a potential well''.  Instead the soliton behaves more like a
classical particle encountering a potential barrier.  Broderick and de Sterke
conjecture that if a defect supports bound states, then a particle approaching
it should ``see'' a potential well, and if it supports no bound state, then an
approaching gap soliton should see a barrier~\cite{BD:98}. Our numerical
simulations and theory, based on energy conservation and resonant energy
transfer illustrate that the situation is more complex.  Indeed, both the
``potential well like'' and ``potential barrier like'' behavior are possible
for a defect that supports bound states.

The reflection of these gap solitons is well-explained by
Figure~\ref{fig:I_v_freq}.  To test our hypothesis, we post-process the
numerical experiment as follows.  At each time step, we compute the projection
of the solution onto the linear defect mode of this defect, and find that when
the gap soliton is directly over the defect, the projection onto the bound
state accounts for only $6\%$ of the total $L^2$ norm of the solution, and
after the soliton escapes, the projection accounts for less than $0.2\%$ of
the solution.  As there is no resonant exchange of energy between soliton and
defect mode, the soliton escapes.

\begin{figure}
\begin{center}
\includegraphics[width=\figwidth]{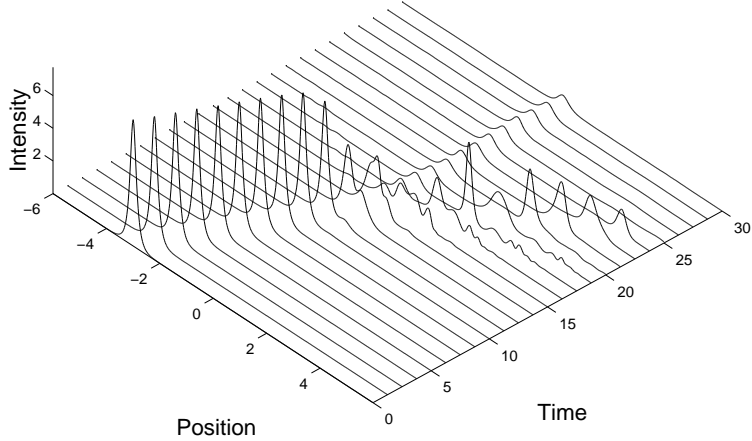}
\caption{(Experiment~1.2) With $\delta=2$, a defect mode of significant
intensity remains behind after the soliton passes through. (Due to the
absorbing boundary conditions used in the simulations, the gap soliton
dissipates as it approaches the edge of the computational domain)}
\label{fig:partcapture}
\end{center}
\end{figure}

\nit \textbf{Experiment~1.2} (Trapping for larger intensities)
Figure~\ref{fig:I_v_freq} suggests that gap solitons with larger values of the
detuning $\d$, may interact more strongly with the defect.  We therefore run
the experiment again with $\delta = 2$ (frequency $\w =-1.72$) below
$\w_0$, and $v=0.2$ the results of which are shown in
Figure~\ref{fig:partcapture}.  
When the soliton encounters the defect, it seems to split into three parts: a
transmitted soliton, a trapped mode, and radiation.  The mode that remains at
the defect has only about $16\%$ of the total intensity of the incoming gap
soliton.  Remarkably, at the end of the computation, the captured state's
frequency is approximately $\w = -1.7$ and the solution's total intensity is
such that the trapped energy is described by a point very close to the
nonlinear bound state curve of Figure~\ref{fig:I_v_freq}. This supports the
first part of our hypothesis. For small values of $v$, the amplitude and
frequency of the trapped state do not seem to depend on $v$.  Above a certain
larger velocity, significantly less energy is trapped by the defect,
suggesting that the hypothesis needs refinement for large velocities.

\begin{figure}
\begin{center}
\includegraphics[width=\figwidth]{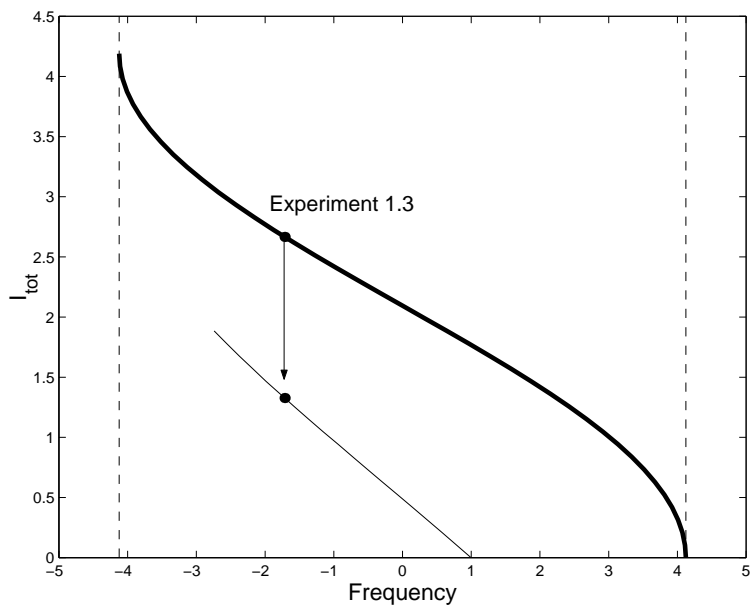}
\caption{Intensity vs.\ frequency for the gap soliton (bold) and a
nonlinear mode for defect with parameters $(\w_0,k)=(1,4)$. Although the
defect is the same as that used in Figure~\ref{fig:I_v_freq} excepting a sign
in the definition of $V(Z)$, the defect mode curve is further to the right and
closer to the gap soliton curve, predicting greatly improved trapping.}
\label{fig:I_v_freq2}
\end{center}
\end{figure}

\nit \textbf{Experiment~1.3} (Refining these results) Clearly, this
computation shows that we can use a defect to trap a significant portion of
the electromagnetic energy in a gap soliton.  However to this point, the gap
solitons we have captured have had very high intensities, and further, only a
small amount of the solitons energy is trapped by the defect.  It is of
interest to trap lower intensity pulses, and it would be preferable if a
larger fraction of the soliton's energy were trapped by the defect. The
nonlinear defect modes always bifurcate to the left from the linear defect
mode frequency for increasing intensity. Consider the defect defined
by~\eqref{eq:defectsolve} with $k=4$ as before, but now letting $\w_0=1$.  This
leaves $\k(Z)$ unchanged, while changing the sign of $V$.  This moves the base
of the nonlinear bound state curve to $\w_0 = +1$, so that the
intensity--frequency curves for the nonlinear defect mode and the gap solitons
are significantly closer together; see Figure~\ref{fig:I_v_freq2}.  
If we examine the interaction of $\d=\pi/2$ ($\w=0$) gap solitons with each of
these defects, the bifurcation diagrams anticipate that the $\w_0=1$ defect
will capture a lot of energy from the pulse, while the $\w_0=-1$ defect will
reflect or transmit the pulse, depending on its incoming velocity. Numerical
experiments show this to be the case.

We can further improve trapping using the dark-soliton family of defects by
increasing the ratio $\w_0/\k_\infty$. The gap edge is at $\k_\infty=
\sqrt{\w_0^2+k^2}$, while the defect mode curve starts at $\w_0$ and goes left
with increasing intensity.  The problem with this approach is that as that as
$\w_0 \to \k_\infty$, the the width of the gap increases without bound, while
its depth goes to zero. 

\subsection{Experiment~2: Gap soliton incident on grating defects supporting
multiple bound states} 
\label{sec:general_gratings}

\begin{figure}
\begin{center}
\includegraphics[width=\figwidth]{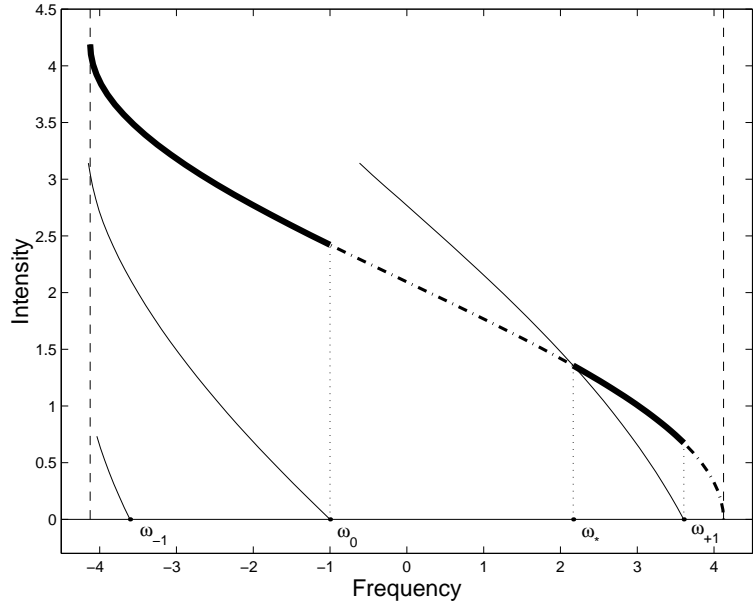}
\caption{Intensity vs.\ frequency for the gap soliton (bold) and the three
nonlinear mode for defect with parameters $(\w_0,k,n)=(-1,2,2)$.  Trapping is
possible for frequencies on the thickened section of the gap soliton curve.}
\label{fig:trap_heuristic}
\end{center}
\end{figure}

We now show how to use the generalized dark soliton defects of
Section~\ref{sec:more_genl} to more efficiently capture solitons. As pointed
out in the paragraph following~\eqref{eq:gapwidth}, we can use defects given
by~\eqref{eq:gen_defect} to fix the spectral gap and study the interaction of
gap solitons with defects of different widths.  We next study gap solitons
incident on a grating of this form with $(\w_0,k,n)=(-1,2,2)$.  This defect is
twice the width of the dark soliton defect grating of
Section~\ref{sec:dsgrating}, but has the same limiting profile far from the
defect region.  From, Appendix~\ref{sec:general_defect}, the defect supports
three linear bound states, with ground state frequency $\w_0=-1$ and excited
states $\w_{\pm 1}=\pm \sqrt{13}$. Branches of nonlinear defect modes
bifurcate from each of these linear modes.  Figure~\ref{fig:trap_heuristic} is
the analog of Figure~\ref{fig:I_v_freq} for this defect.  To the left of the
indicated frequency $\w_*$, the $\w_{+1}$-nonlinear defect mode has greater
intensity than the gap soliton. The frequencies $\w_{-1},\ \w_0,\ \w_*$, and
$\w_{+1}$ divide the band gap into 5 regions.  In the regions $-\k_\infty < \w
< \w_0$ and $\w_* <\w < \w_{+1}$ we expect by the same mechanisms as in
Experiment~1.2 to find trapping of energy, while in the regions $\w_0< \w <
\w_*$ and $\w_{+1}< \w < \k_\infty$, we do not expect trapping.  In the
segment $-\k_{\infty} < \w < \w_{-1}$, we expect complex behavior because two
trapped nonlinear modes coexist.  We expect the most efficient capture for
solitons with frequency slightly greater than $\w_*$, for which a nonlinear
bound state exists of slightly lower total intensity than the incoming pulse.

\noindent {\bf Experiment~2.1:} (The trapping region $\w_* < \w < \w_{+1}$) 

\begin{figure}
\begin{center}
\includegraphics[width=\figwidth]{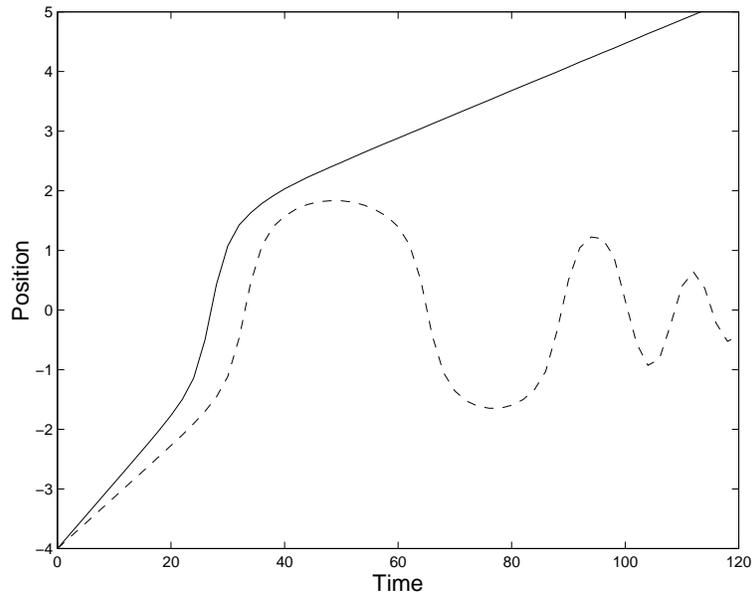}
\caption{(Experiment~2.1a) The position vs.\ time of an escaping and a
captured gap soliton.  Note that the instantaneous velocity increases when the
gap soliton is in the defect region ($Z$ near zero). $\d=0.9$, $v$ near $v_c
\approx 0.103$, the potential as described in
subsection~\ref{sec:general_gratings}.}
\label{fig:capt_escape}
\end{center}
\end{figure}

\begin{figure}
\begin{center}
\includegraphics[width=\figwidth]{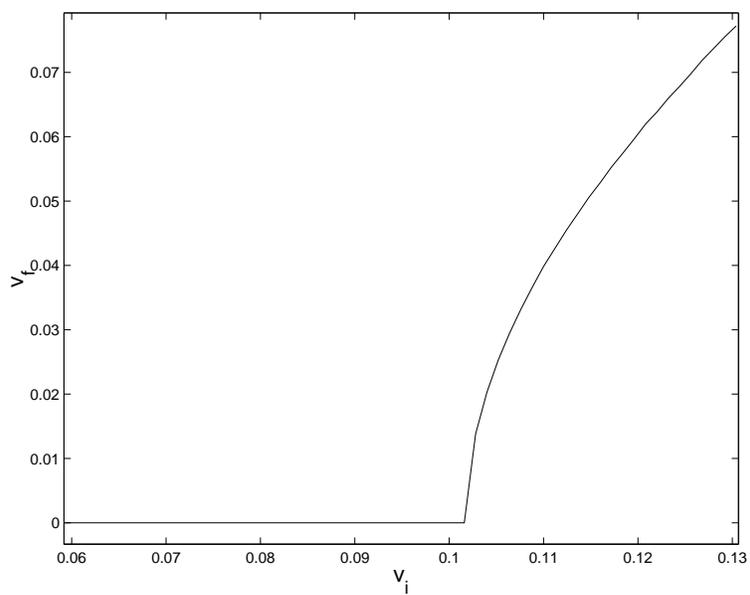}
\caption{(Experiment~2.1a) Initial velocity ($v_i$) vs.\ final velocity ($v_f$)
of gap soliton, parameters as in~\ref{fig:capt_escape} with variable $v$.}
\label{fig:vi_vf}
\end{center}
\end{figure}

\noindent \textbf{(2.1a)} We first examine gap solitons with $\d=0.9$, ($\w
\approx 2.6$) which lie just to the right of $\w_*$ in
Figure~\ref{fig:trap_heuristic}. Trapping here is relatively efficient,
because a soliton can transfer almost all its amplitude to the nonlinear
defect mode of the same frequency and slightly lower intensity. We find
trapping for gap solitons slower than a critical velocity of about
$v_c=0.102$.  In Figure~\ref{fig:gscapture}, we show the evolution of a gap
soliton, initially propagating to the right, which gets trapped.  In
Figure~\ref{fig:capt_escape}, we show the position vs.\ time plot for a gap
soliton that gets trapped and one that escapes. In both cases, the gap soliton
speeds up on reaching the defect, consistent with the defect acting as a
potential well. Figure~\ref{fig:vi_vf}, shows the output soliton velocity as a
function of the soliton input velocity in a region near the critical
velocity. The figure indicates a sharp transition at a critical velocity from
gap solitons which are trapped to gap solitons which propagate through.
\begin{figure}
\begin{center}
\caption{(Experiment~2.1a) A typical picture of the capture of a gap soliton by
a defect centered at $Z=0$.}
\label{fig:gscapture}
\end{center}
\end{figure}

\begin{figure}
\begin{center}
\includegraphics[width=\figwidth]{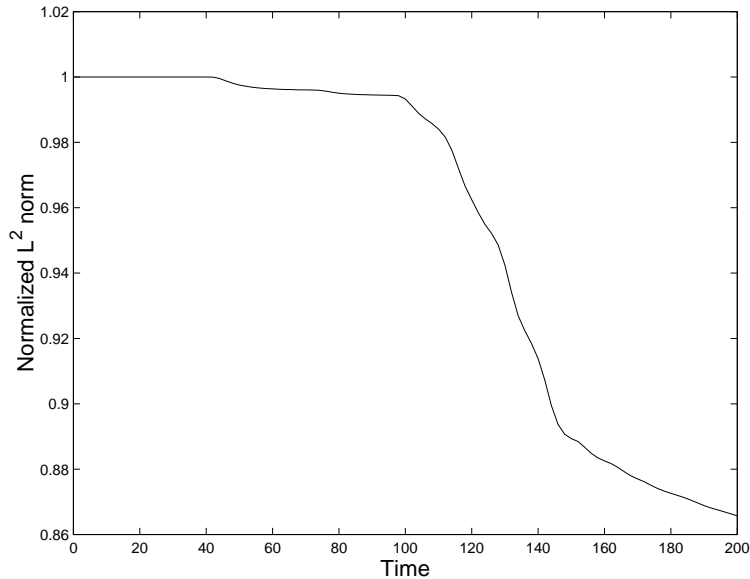}
\caption{Decay of the local $L^2$ norm (normalized) for Experiment~2.1a.}
\label{fig:L2norm}
\end{center}
\end{figure}

Although the NLCME system~\eqref{eq:nlcme_nondim} conserves $I_{\rm tot}$,
radiation may carry some energy away from the defect.  The computations are
performed on a finite domain with absorbing boundary conditions, so that
radiation losses can be measured by monitoring the local $L^2$ norm,
\textit{i.e.}
$$
\text{Local $L^2$ norm } = \left(\int_D \abs{\Eplus}^2 +\abs{\Eminus}^2
 dZ \right)^{\frac{1}{2}},
$$
where $D$ is a bounded region containing the defect; see
Figure~\ref{fig:L2norm}. By time $t=120$, the energy has been transferred from
the soliton to the nonlinear defect mode, but the system continues losing
energy to radiation at a constant rate for the length of the simulation.

\begin{figure}
\begin{center}
\includegraphics[width=\figwidth]{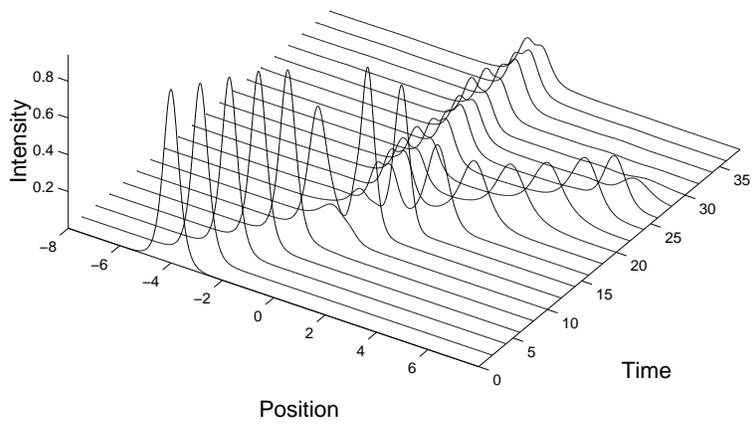}
\caption{(Experiment~2.1b) Partial capture; an incident gap soliton results in
part of its energy captured in the defect and part transmitted as a lower
energy gap soliton}
\label{fig:capture_part}
\end{center}
\end{figure}

\noindent\textbf{(2.1b)}For the slightly a smaller value of $\d=0.6$, the
distance between the gap soliton curve and the nearby defect-mode curve in
Figure~\ref{fig:trap_heuristic} is increased.  Some of the energy is deposited
in a defect mode, while the remaining energy appears to propagate as a
diminished gap soliton plus small radiation; see Figure~\ref{fig:capture_part}.

There is not space to report in detail on all of the behaviors found in the
numerical simulations for this defect.  As expected, gap solitons with
frequency $\w< \w_0$ are trapped in a similar manner to those in
Experiment~1.2. When $\w < \w_{-1}$, then all three nonlinear modes are
excited by the gap soliton.  More unexpectedly, in the region $\w_0 < \w <
\w_*$, the gap soliton, while never captured, is never reflected either.  In
this frequency range, for every initial velocity as low as $v=.0006$, the
soliton is transmitted after slowing down slightly when encountering the
defect.

\noindent{\bf Experiment~2.2:} (Wider defects)

\begin{figure}
\begin{center}
\includegraphics[width=\halfwidth]{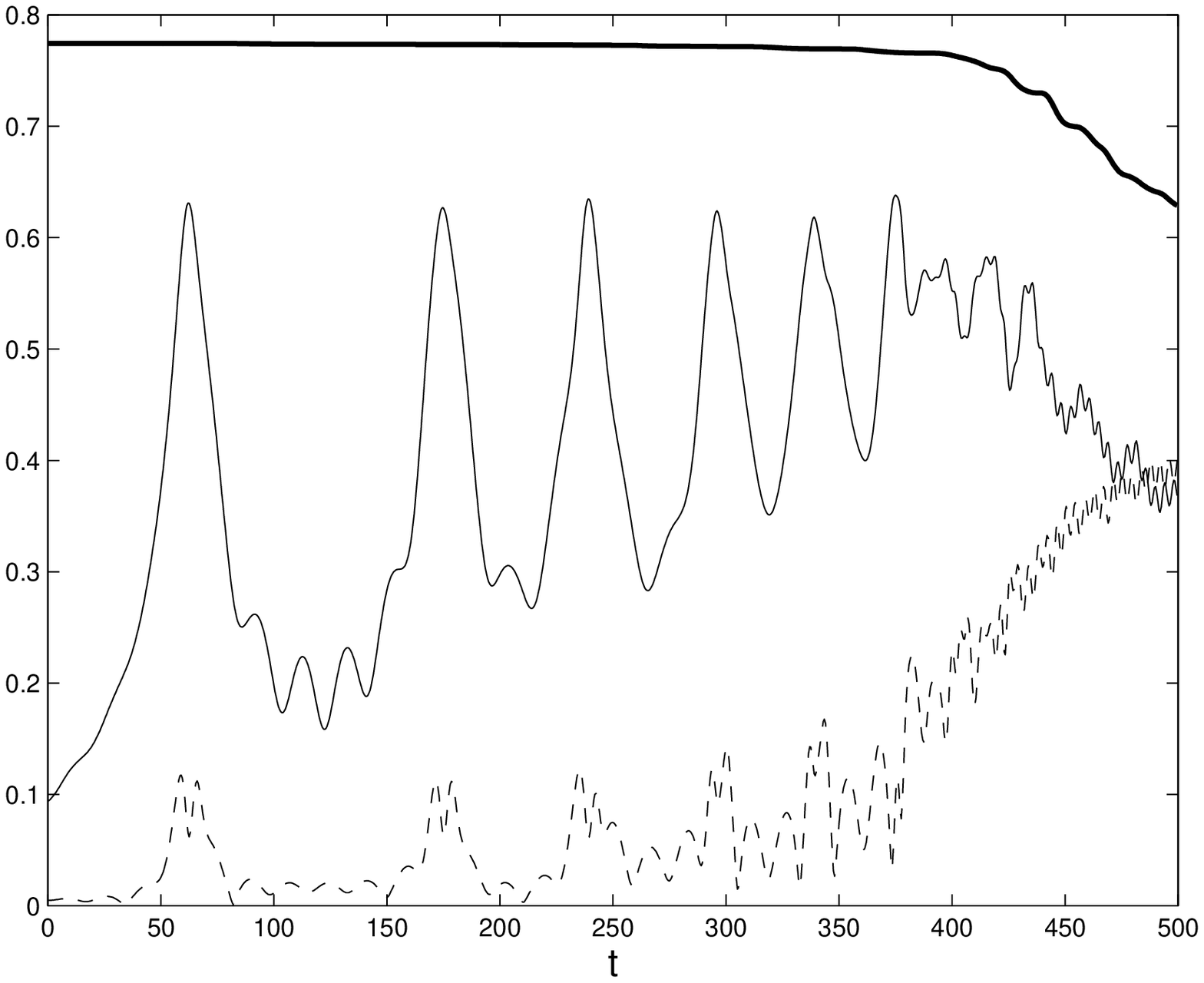}
\includegraphics[width=\halfwidth]{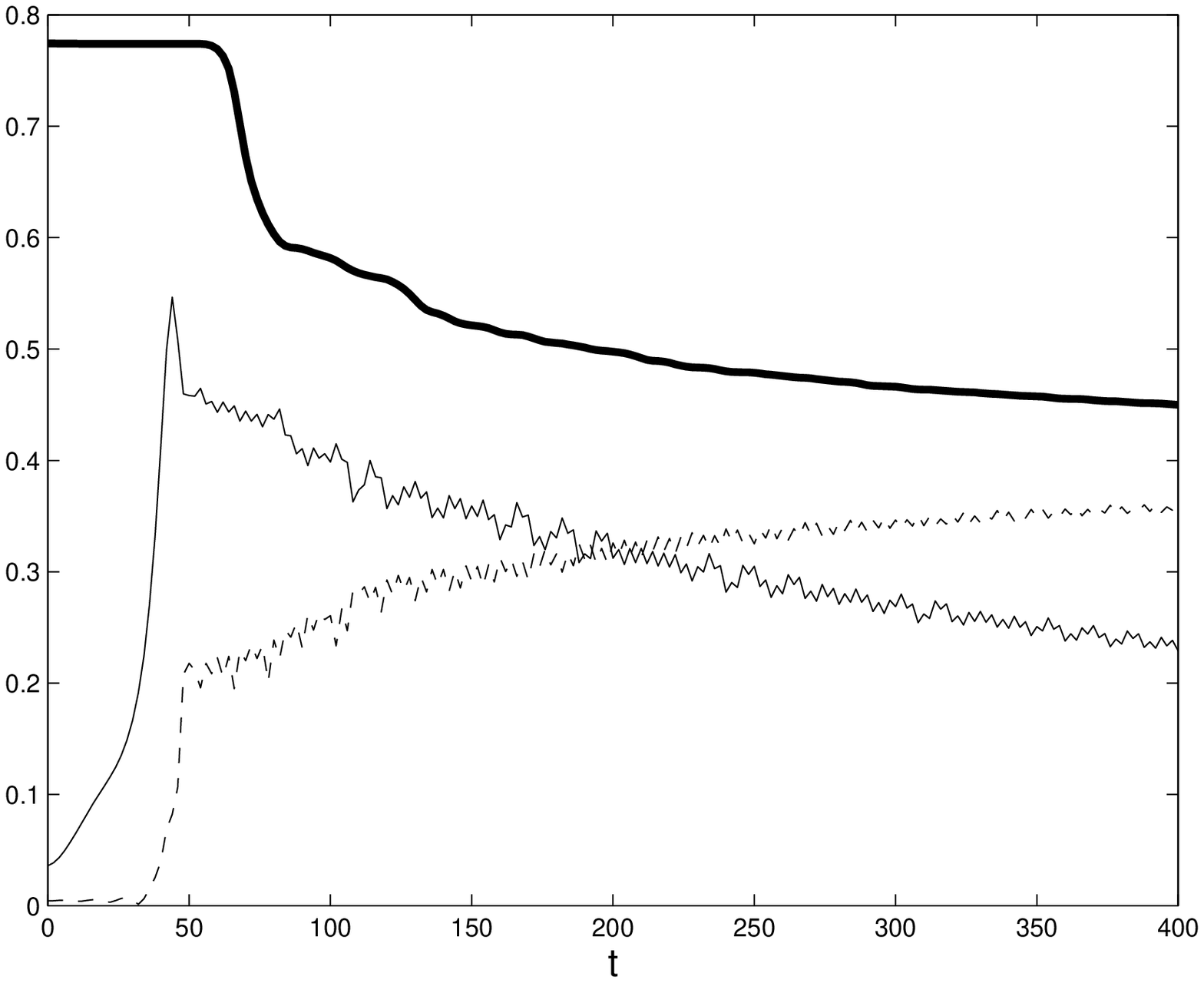}
\caption{Local $L^2$ norm of the solution (bold), the projection onto the
$\w_{+2}$ eigenmode (solid) and the $\w_{+1}$ mode (dashed) for
Experiment~2.2a (left) and 2.2b (right).}
\label{fig:projections}
\end{center}
\end{figure}

\noindent{\bf (2.2a):} By widening the defect, we may place more eigenvalues
closer to the edges of the band gap which might then be used to trap gap
solitons with even smaller $\d$.  Using a defect with parameters $(\w_0,k,n) =
(-1, 1.6,2.5)$ (keeping $\k_\infty = \sqrt{17}$ as in previous sections),
which has 5 eigenvalues $\w_0 = -1$, $\w_{\pm 1}= \pm\sqrt{281}/5$, $\w_{\pm
2} \pm \sqrt{409}/5$, we captured a soliton with~$\d=0.45$, although with the
velocity significantly reduced to about $v=0.025$.  We found that the defect
with $k=n=2$ described in Experiment~2.1 reflects this gap soliton, as its
central frequency is to the right of the defect mode curve in
Figure~\ref{fig:trap_heuristic}. The dynamics of the soliton captured by the
present defect is shown in Figure~\ref{fig:smalldeltacapture}. At about
$t=400$ the trapped mode begins to lose intensity to radiation, as is more
clearly seen in Figure~\ref{fig:projections}a; note the decay of the local
$L^2$ norm beginning around $t=400$.
\begin{figure}
\begin{center}
\caption{(Experiment~2.2a) Capture of a gap soliton with $\d=0.45$ by a wide
well.}  
\label{fig:smalldeltacapture}
\end{center}
\end{figure}

Also shown in Figure~\ref{fig:projections} are the numerical projection
onto the linear eigenfunctions belonging to $\w_{+2}$ and $\w_{+1}$. At
capture time ($t \approx 60$), the solution is dominated by the $\w_{+2}$
mode.  At longer times, however, this trapped mode is not persistent and the
energy is transferred from $\w_{+2}$ to $\w_{+1}$, with a lot of energy lost
to radiation and very little in the other three eigenmodes.

\noindent{\bf(2.2b):} The effect of defect width can be further explored.  In
the next experiment, we widen the defect further, by choosing parameters
$k=4/3$, $n=3$ in~\eqref{eq:genkappa}, and choose the same gap soliton
parameters as in the above paragraph.  The defect is slightly wider than in
Experiment~2.2a, though it still supports five linear bound states, with
eigenvalues slightly smaller than in the previous paragraph: $\w_0 = -1$,
$\w_{\pm 1} \pm \sqrt{89}/3$, $\w_{\pm 2}= \pm \sqrt{137}/3$. In this case,
the trapping is significantly less effective. A much smaller bound state is
trapped.  As in the previous experiment, most of the energy is localized in
the modes belonging to $\w_{+1}$ and $\w_{+2}$, although, as seen in
Figure~\ref{fig:projections}, the $\w_{+1}$-mode begins growing sooner.

\subsection{Experiment~3: Defect arrays}
\begin{figure}
\begin{center}
\includegraphics[width=\figwidth]{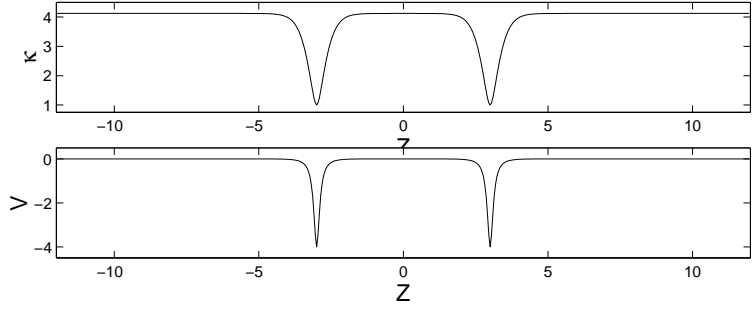}
\caption{(Experiment~3) $\k(Z)$ and $V(Z)$ for an array of two defects.}
\label{fig:2defect}
\end{center}
\end{figure}

\begin{figure}
\begin{center}
\includegraphics[width=\figwidth]{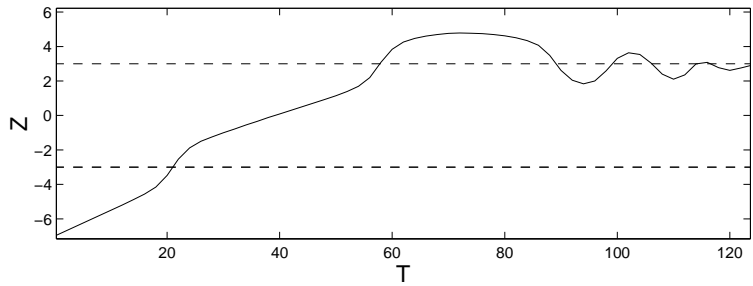}
\caption{(Experiment~3) Position vs.\ time for a gap soliton incident on
an array of defects. Defect positions given by dashed lines.}
\label{fig:2well_ZvsT}
\end{center}
\end{figure}

Figure~\ref{fig:vi_vf} shows that gap solitons which are not trapped by a
defect may be severely slowed.  This suggests that an array of
defects of the type discussed above can be used to successively slow and then
trap a gap soliton.  Using a pair of defects, we have been able to trap a gap
soliton whose initial velocity was 50\% higher than the critical velocity
found in Experiment~2.1.  We construct defect arrays
$\k_2(Z)$ and $V_2(Z)$ by forming $\k(Z)$ and $V(Z)$ of
subsection~\ref{sec:general_gratings} and then letting $\k_2 = \k(Z-Z_1) +
\k(Z-Z_2)- \k_\infty$ and $V_2(Z) = V(Z-Z_1) + V(Z-Z_2)$.  Such gratings with
$Z_1=-3$ and $Z_1=3$ are shown in Figure~\ref{fig:2defect}.  In
Figure~\ref{fig:2well_ZvsT} we show the position vs.\ time for gap soliton
with $\d=.9$, $v=.15$, which is slowed by the first well and then captured by
the second.

\subsection{Experiment~4: Side barriers}
\label{sec:sidebarriers}
\begin{figure}
\begin{center}
\includegraphics[width=\figwidth]{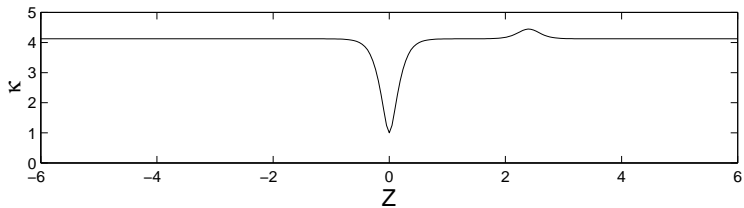}
\caption{(Experiment~4) The modified defect described
in~\ref{sec:sidebarriers}.} 
\label{fig:bumpright}
\end{center}
\end{figure}

\begin{figure}
\begin{center}
\includegraphics[width=\figwidth]{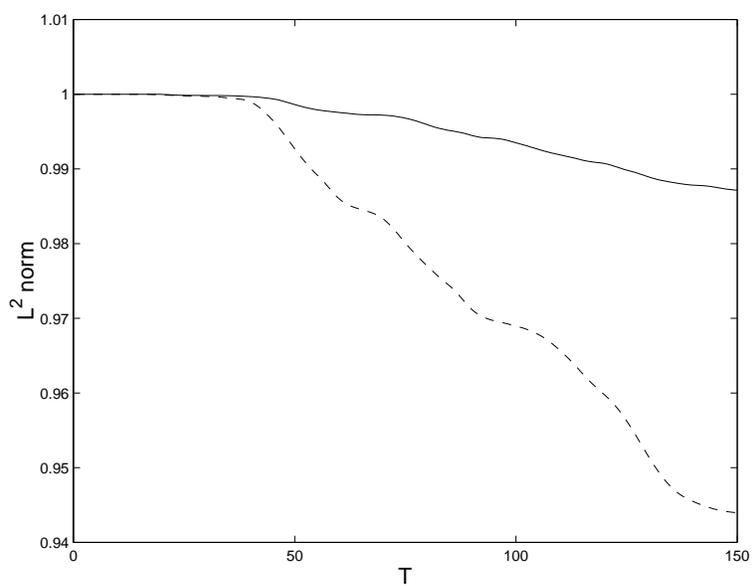}
\caption{(Experiment~4) Local $L^2$ norm as a function of time for gap soliton
with $v=.2625$ (solid) and $v=.3$ (dashed).}
\label{fig:bumprightL2}
\end{center}
\end{figure}

In the previous sections, we captured light by transferring energy from a gap
soliton to a nonlinear defect mode.  In this section, by contrast, we trap a
moving soliton between two obstacles. In Experiment~1.1, solitons with
frequency to the right of the ground state frequency were slowed but not
trapped upon encountering the defect. We modify the defect by adding a bump or
``potential barrier'' away from the main defect; see
Figure~\ref{fig:bumpright}.
This configuration of defects traps the gap soliton in a novel way.  Instead
of a bound state forming near the minimum of $\k(Z)$, the gap soliton bounces
back and forth between the old ``well'' and the new ``bump'' that has been
added.  Further, it captures a pulse with an incident speed of $v=.3$, about 3
times the critical velocity for the generalized dark soliton grating described
in Experiment~2.1.  In addition, the rate of energy loss for gap solitons
captured by this 
defect is significantly reduced.  Figure~\ref{fig:bumprightL2} shows the rate
of energy loss for gap solitons with $\d=.9$ and velocities $v=.2625$ and
$v=.3$.  It shows that, although it can capture gap solitons moving this fast,
as the speed increases, the efficiency of the capture decreases.

\section{Nonlinear damping effects} 
\label{sec:nldamping}
As we have shown in the preceding section, the gap solitons for which we have
been able to find interesting capture behavior have all had large values of
$\d$. (So large, in fact, that perhaps silica fibers would not be able to
support pulses of that intensity.)  One potential way to make use of our
theoretical solutions would be to use fibers with larger nonlinear refractive
index $n_2$.  We see by equation~\eqref{eq:gs_scales} that the intensity of a
gap soliton, for fixed $v$ and $\d$, is inversely proportional to $n_2$.
\textit{Chalcogenide} fibers, for which $n_2$ is as much as 500 times larger
than in silica fibers, are a promising material in which one could potentially
observe the phenomena discussed above at lower intensities.  Unfortunately, in
chalcogenide fibers, the imaginary part of $n_2$ is also significantly larger,
corresponding to nonlinear damping arising from \textit{multiphoton
absorption}. By chooseing a chalcogenide glass composition that minimizes both
two photon and three photon absorption, one can achieve an $n_2$ nearly 500
times silica while suffereing a multiphoton loss of of a few percent at
intensitiess requeired for a nonlinear phase shift of $\pi$.

In terms of the coupled mode equations, a complex cubic refractive index gives
rise to a complex coefficient $\G$.  Due to the symmetries of the NLCME
and the gap soliton, the magnitude of $\G$ is unimportant, as gap solitons
have intensity that scales as $1/\G$.  What will be important is the ratio
$\G_i/\G_r$. In this case we are more interested in simply simulating
the propagation of pulses which at $t=0$ correspond to gap solitons.  The
strength of the damping is proportional to $I_{\rm max}$, or $\sin^2
\tfrac{\d}{2}$. 
Therefore as we decrease $\d$ the effect of the nonlinear damping should 
be decreased.  However, if decreasing $\d$ requires a decrease in the gap
soliton velocity $v$ for trapping, so that pulses will have more time
to decay as they propagate before reaching the defect.  We ran one set of
experiments with $\G_i/\G_r = 0.1,\ 0.01,\ \text{and}\ 0.001$, with
$v=0.2$ and $\d=0.9$, and with the generalized dark soliton grating with
$\w_0=-1$, $k=2$, and $n=2$.  Without nonlinear damping, the defect will not
capture these solitons. The soliton was initialized $5$ units to the left of
the defect center.  With the damping ratio $.1$, the gap soliton is mainly
damped before it even reaches the defect.  With ratio $.01$, the pulse loses
just enough energy that much of it is captured upon reaching the defect.  With
the ratio $.001$, the gap soliton propagates through the defect untrapped.

\section{Summary and discussion}
\label{sec:summary}
Gap solitons are localized nonlinear bound states which propagate in periodic
structures.  We have investigated by analytical and numerical methods the
possibility of capture of gap solitons by the introduction of appropriately
designed defects, spatially localized deviations from exact periodicity.

We first displayed interesting classes of defects which support trapped defect
modes of the linear coupled mode equations. We then showed that these linear
defect modes, deform into \textit{nonlinear defect modes} of the nonlinear
coupled mode equations.  Bifurcation diagrams of total intensity ($I_{\rm
tot}$) versus frequency suggest the following hypothesis: for sufficiently low
velocities, a gap soliton incident on a defect will transfer its energy to a
nonlinear defect mode localized at the defect provided there is one of the
same frequency (resonance) and lesser total intensity (energetically
accessible).  This hypothesis is supported by an extensive series of numerical
investigations.  An understanding of the dependence of the critical velocity
requires further investigation.

We have studied the interaction of gap solitons with defects which support one
or multiple linear (and therefore nonlinear) defect modes. In
Section~\ref{sec:general_gratings} we show how secondary defect
modes can be used to trap energy from lower-amplitude gap solitons, which
correspond more closely to the regime accessed thus far in physical
experiments. However, in contrast to trapping by a defect mode with a single
defect, when gap soliton energy is transferred to the modes of a multimode
defect, the dynamics of the localized energy is quite complicated.

We believe that a finite-dimensional model incorporating both soliton and
defect mode degrees of freedom could be very useful in understanding the
capture problem. In these models the soliton is modeled by several parameters
(\textit{e.g.}, position, width, phase) and the defect mode by its
"amplitude". A system of ordinary differential equations approximating the
dynamics of a soliton interacting with a defect can be obtained from an
\textit{effective Lagrangian}, which is a function of these collective
variables. Broderick and de Sterke~\cite{BD:98} have studied such a model
which does not take into account degrees of freedom 
available in the defect modes.  Their model displays some of the observed
behaviors but, as we have seen, the mechanism of resonant energy transfer must
be included to provide a full explanation.  

Similar finite-dimensional models have been studied in~\cite{FKV:92a, FKV:92,
FPM:94} for the sine-Gordon, $\phi^4$ and nonlinear Schr\"odinger equations.
Comparison of models for the trapping of sine-Gordon kinks with and without
the defect-mode degrees of freedom~\cite{M:85, FKV:92} demonstrates the
necessity of allowing the additional modes of oscillation.  We have applied
tools of dynamical systems theory to similar reduced models for soliton like
structures of the sine-Gordon equeation~\cite{GHW:01} and
NLS~\cite{GHW:01a}. These studies give insight into the nature of the set of
states incident on the defect resulting in transmission without capture,
capture, and capture for all time.  Closer qualitative agreement with the full
dynamics is obtained by inclusion of a damping term, reflecting the coupling
to radiation modes.

\appendix
\section{Calculation of gap soliton characteristics}
\label{sec:gscalc}

In this appendix, we derive dimensional values of $\k_\infty$ and $\Tilde\G$
in equations~\eqref{eq:K_def} and~\eqref{eq:G_def}. We use these quantities to
estimate the physical parameter values corresponding to the simulations
performed in Section~\ref{sec:simulations}.

\subsection{Dimensional values of $\tilde \k_\infty$ and $\tilde \Gamma$}
It is common to work with the cubic refractive index~\cite[p.40, 582]{A:95},
$n_2$ (also denoted $n_2^I$): 
\begin{equation}
n_2 = 3\chi^{(3)}/4\eps_0 c \Bar n^2,\label{eq:n2I}
\end{equation}
quoted in units of ${\rm m}^2/{\rm W}$. This  gives 
\begin{equation}
\Tilde \G = \frac{4 \pi \eps_0 c \Bar n n_2}{\lambda_B}.
\label{eq:Gammatilde}
\end{equation}

We list the parameters needed, as obtained from Eggleton, \textit{et.\
al.}~\cite{E:97} and from standard sources:
\begin{equation}
\label{eq:exp_param}
\begin{split}
\lambda_B &= 1053\, {\rm nm}; \\
\Bar n &= 1.45; \\
\Delta n &= 3 \times 10^{-4}; \\
n_2 &= 2.3 \times 10^{-20} \, {\rm m^2} / {\rm W}; \\
c &= 2.98 \times 10^{8}\, {\rm m}/{\rm s} ; \\
\eps_0 &= 8.85 \times 10^{-12}\, {\rm C}/{\rm Nm}^2.
\end{split}
\end{equation}
Then the wavenumber in the medium is given by
\begin{equation}
k_B = 2 \pi \frac{\Bar n}{\lambda_B} = 8.7 \times 10^6 \, {\rm m}^{-1},
\end{equation}
and ${\Tilde \k}_\infty$ (see~\eqref{eq:K_def}) and $\Tilde \G$ by%
\begin{align}
 {\Tilde \k}_\infty &= 900 \, {\rm m}^{-1}, \,\ {\rm and}\ \\
 \Tilde \G &= 1.06 \times 10^{-15} \, 
\frac{{\rm C}^2}{{\rm N}^2 {\rm m}}.
\end{align}%

\subsection{Converting from nondimensional to dimensional form}
Inverting the nondimensionalization relations~\eqref{eq:nd_def}, the
dimensional length, time and electric field scales can be obtained from the
following relationship
\begin{subequations}
\begin{align}
\cZ &= \frac{\k_\infty}{{\Tilde \k}_\infty} =
 \frac{\k_\infty\lambda_B}{\pi \Delta n},\;
\cT = \frac{\k_\infty\lambda_B}{\pi \Delta n}\frac{\Bar n}{c}
\label{eq:defZ}\\
\cE^2 &= \frac{\G \Tilde \k_\infty }{\Tilde \G \k_\infty} =
 \frac{\G \Delta n}{4 \k_\infty \eps_0 c \Bar n n_2},
\label{eq:defEscale}
\end{align}
\label{eq:SCALES}	
\end{subequations}
where we have also used~\eqref{eq:K_def} and~\eqref{eq:G_def}.

\subsection{Dimensional Experimental Parameters}
In terms of the solution to the NLCME, the (scalar, dimensional) electric
field is given by 
$$
E = \eplus e^{ik_B(z-ct/\Bar n)} + \eminus e^{-ik_B(z+ct/\Bar n)} + cc
$$
where $cc$ represents the complex conjugate.  The mean amplitude of the
electric field is thus given by
$$
\abs{E}^2 =2\left( \abs{\eplus}^2 + \abs{\eminus}^2\right) 
= 2 \cE^2\left( \abs{\Eplus}^2 + \abs{\Eminus}^2\right),
$$
neglecting phase and cross terms. The maximum intensity~\cite{Agrawal} is 
\begin{equation}
 I = \frac{1}{2} \eps_0 c \Bar n \max\abs{E}^2 
= \eps_0 c \Bar n \cE^2 I_{\rm max}.
\end{equation}
Note that this is a scaling of the nondimensional quantity $I_{\rm max}$
defined in~\eqref{eq:Imax}. Combining this with~\eqref{eq:Imax}
and~\eqref{eq:defEscale}, 
\begin{equation}
I = \frac{2\Delta n \sqrt{1-v^2}}{n_2 (3-v^2)} \sin^2 {\frac{\d}{2}}
\label{eq:gs_scales}
\end{equation}
Scaling~\eqref{eq:FWHM} by~\eqref{eq:defZ} and using that $\k_\infty =
\sqrt{\w^2 + n^2 k^2}$ gives
$$
\text{Dimensional} \, {\rm FWHM} =  
\frac{\lambda_B}{\pi \Delta n} \cdot
\frac{2\sqrt{1-v^2}}{\sin\d} 
\cosh^{-1}\sqrt{1+\cos^2{\frac{\d}{2}}}. 
$$
The temporal width is just 
$$
\text{Dimensional} \, {\rm FWHM}_{\rm temporal} = 
\frac{\Bar n \lambda_B}{\pi c \Delta n} \cdot
\frac{2\sqrt{1-v^2}}{v \sin\d} 
\cosh^{-1}\sqrt{1+\cos^2{\frac{\d}{2}}}. 
$$
The coupling function $\Tilde\k(z)$ has units of inverse length,
therefore, the dimensional equivalent of~\eqref{eq:depth} scaled
by~\eqref{eq:defZ} is
$$
\Delta_* =  \frac{\pi \Delta n}{\lambda_B}
\left(1 - \frac{\abs{\w}}{\sqrt{\w^2 + n^2k^2}}\right).
$$
The defect width~\eqref{eq:defect_fwhm} scaled by~\eqref{eq:defZ} is
$$
{\rm FWHM} = \frac{\lambda_B}{\pi \Delta n} \cdot
\frac{2\sqrt{\w^2+k^2n^2}}{k}
\tanh^{-1} {\left(
\frac{\sqrt{2\abs{\w}\sqrt{\w^2+n^2k^2}+n^2k^2-2\w^2}}{2nk} 
\right)}.
$$

With this information, we can construct the dimensional parameters describing
the numerical experiments in Section~\ref{sec:simulations}.
Table~\ref{tab:defects} describes the defects and Table~\ref{tab:pulses}
describes the gap solitons. For simplicity, all pulse measurements are
computed for $v=0$ so the spatial width must be used.

\begin{table}[hbt]
\begin{center}
\begin{tabular}{|r||r|r|r||c|c|} \hline
\multicolumn{1}{|c||} {Experiment}&
\multicolumn{1}{|c|} {$\w$}&
\multicolumn{1}{|c|} {$k$}&
\multicolumn{1}{|c||} {$n$}&
\multicolumn{1}{|c|} {$\Delta_*$ (${\rm m}^{-1}$)}&
\multicolumn{1}{|c|} {Defect FWHM (mm) }\\ \hline
1 & {$\pm 1$} & 4 & 1 & 680 & 1.6\\
2.1 & -1 & 2 & 2 & 680 & 3.1 \\
2.2a & -1 & 1.6 & 2.5 & 680 & 3.9 \\
2.2b & -1 & 1.33 & 3 & 680 & 4.7\\
\hline
\end{tabular}
\end{center}
\caption{Experimental parameters describing the defects in numerical
simulations.} 
\label{tab:defects}
\end{table}

\begin{table}[hbt]
\begin{center}
\begin{tabular}{|r||r||c|c|} \hline
\multicolumn{1}{|c||} {Experiment}&
\multicolumn{1}{|c||} {$\d$}&
\multicolumn{1}{|c|} {$I$ (${\rm GW}\cdot{\rm cm}^{-2}$)}&
\multicolumn{1}{|c|} {Gap Soliton FWHM (mm) }\\ \hline
1.1 & 0.9 & 490 & 2.3 \\
1.2 & 2 & 1800 & 1.3 \\ 
1.3 & $\pi/2$ & 1300 & 1.5 \\
2.1a & 0.9 & 490 & 2.3 \\
2.1b & 0.6 & 230 & 3.4 \\
2.2 & 0.45 & 130 & 4.4 \\
\hline
\end{tabular}
\end{center}
\caption{Experimental parameters describing the gap solitons in numerical
simulations.} 
\label{tab:pulses}

\end{table}

\section{Defect gratings with a prescribed mode}
\label{sec:general_defect}
In this section, we define a simple procedure for generating grating profiles
with a given band gap, eigenvalue, and an eigenmode of prescribed shape.
We may specify $\w$ and look for solutions of the form
(see~\eqref{eq:Eansatz}) 
\begin{equation}
\binom{\Eplus}{\Eminus} = e^{-i\w T} e^{i\sigma_3\Theta(Z)} f(Z)
\binom{v_+}{v_-}
\label{eq:Epm-mode}
\end{equation}
where $v_\pm$ are constants, $\partial_Z \Theta(Z) =V(Z)$, and $f(Z)$ is a
real scalar function such that $f(Z) \sim e^{-k\abs{Z}}$ as 
$\abs{Z} \to \infty$.  

Then
\begin{equation*}
\pdiff{\Eplus}{Z} =  e^{-i\w T}\left( f'+iVf \right) e^{i\Theta(Z)} v_+.
\end{equation*}
Letting $g = \frac{f'}{f},$ this may be rewritten
\begin{equation*}
\pdiff{E_{\pm}}{Z} = \left( g \pm iV \right) \Eplus.
\end{equation*}
Similarly
Then~\eqref{eq:lcme} becomes
$$
(\w\pm ig) E_{\pm} + \k E_{\mp} =0; 
$$
By~\eqref{eq:Epm-mode},
\begin{equation*}
{\mathcal L} \Vec v =
\begin{bmatrix}  (\w + i g) e^{i \Theta} & \k e^{-i\Theta} \\
 \k e^{i\Theta} &  (\w - i g) e^{-i \Theta}
\end{bmatrix}
\binom{v_+}{v_-} = 0.
\end{equation*}
If~\eqref{eq:Epm-mode} defines a solution of~\eqref{eq:lcme} then
\begin{equation*}
\det {\mathcal L}\ =\ \w^2 + g^2 - \k^2\ =\ 0.
\end{equation*}
Therefore, 
\begin{equation}
\k(Z) = \sqrt{\w^2 + g^2(Z)}. \label{eq:define_kappa}
\end{equation}
Note also that if
$f(Z) \sim e^{-k \abs{Z}}$ 
 then
$\abs{g} \to \pm k \, \text{as} \, Z \to \mp\infty$
and
\begin{equation*}
\k_\infty = \lim_{Z \to \infty} \k(Z) = \sqrt{\w^2 + k^2}.
\end{equation*}
The width of the gap is then equal to $2\k_\infty$.

If $\Vec v$ is a null eigenvector, then
\begin{equation}
\frac {v_-}{v_+} = -e^{2i\Theta} \frac{\w + i g}{\k}. \label{eq:vpm}
\end{equation}
By~\eqref{eq:define_kappa}, we have that
\begin{equation*}
\abs{\frac{\w + i g}{\k}} = 1
\end{equation*}
and therefore
\begin{equation*}
\frac{v_+}{v_-} = e^{i\alpha},\ \alpha\ {\rm real}.
\end{equation*}
Examination of the eventual solution shows that $\alpha$ merely reflects the
invariance of the equations under a constant phase shift.  We therefore set
$\alpha=0$ in the remainder of the argument.  Then, in order to
satisfy~\eqref{eq:vpm}, 
\begin{equation*}
2\Theta = \arg {\frac{-\w + ig }{\k}}.
\end{equation*}
Since $\k$ is positive
\begin{equation}
\Theta =  -\frac{1}{2} \arctan{\frac{g}{\w}}.
\end{equation}
Then
\begin{equation}
V = -\frac{\w g'}{2(\w^2 + g^2)}.
\end{equation}

We may use this method to construct defects that support defect modes of
arbitrary shape with prescribed exponential decay.  If we choose a function
with different exponential decay rates as $Z \to \pm \infty$, then
$\k_{\pm\infty}$ will take two different values.

In the case $\w=0$, we find the same discontinuous limiting behavior as
in~\eqref{eq:zerolimit}. Fortunately, this is merely due to the inadequacy of
the polar decomposition implicit in the definition of $\Theta(Z)$
in~\eqref{eq:Epm-mode}.   For $\w=0$, the entire calculation may be repeated
with $V=0$, and a smooth solution generalizing~\eqref{eq:Fzero} is
generated:
\begin{align}
\k(Z) &= \pm g(Z), \\
v_- &= \mp i v_+.
\end{align}

\subsection*{Example}
\label{sec:generalized}
The defect gratings of Section~\ref{sec:more_genl} which generalize the dark
soliton gratings can be obtained as follows.  Let
$$
f = \sech^n{(kZ)}.
$$
Then
\begin{align}
\k(Z) &= \sqrt{\w^2 + n^2 k^2 \tanh^2 {(kZ)}}; \\
\Theta(Z) &= \frac{1}{2} \arctan{\frac{n k \tanh{(kZ)}}{\w}}; \\
V(Z) &= \frac{\w n k^2 \sech^2{(kZ)}}
               {2(\w^2 + n^2 k^2 \tanh^2{(kZ)})}. 
\end{align}
And
\begin{equation}
\binom{\Eplus}{\Eminus} = e^{-i\w t} e^{\pm \frac{i}{2}
\arctan{\frac{n k\tanh{(kZ)}}{\w}}} \sech^n{(kZ)}.
\end{equation}
Setting $n=1$, we recover the ``dark soliton defect'' of
Section~\ref{sec:DSdefect}.  By varying $n$, while keeping the quantity $\w^2
+ n^2k^2$ fixed, we generate a family of gratings of variable widths with
identical band gaps.

Numerical computations indicate that for $n>1$, the system supports multiple
eigenmodes obeying the following rule.  For $n>0$, the defect supports a total
of $2\lceil n \rceil -1$ eigenmodes where $\lceil n \rceil$ is the smallest
integer greater than or equal to $n$.  The ground state has frequency $\w_0=
\w$ and spatial decay rate $nk$, whereas the excited states occur in pairs
with spatial decay rates given by $(n-j)k$ and frequencies:
\begin{equation}
\w_{\pm j} =\pm \sqrt{\w^2 + (2nj-j^2)k^2}
\label{eq:higherfreqs}
\end{equation}
for all $1 \le j < n$.  It should be possible to derive this formula exactly
using methods of complex analysis developed to study bound states of the
Schr\"odinger equation with potential~\cite{T:62}, as well as expressions for
the associated bound states.

\section*{Acknowledgements} 
RG was supported by an NSF University-Industry cooperative research fellowship
DMS-99-01897. The authors would like to thank Phil Holmes, Vadim Zharnitsky,
Ben Eggleton, and Dmitry Pelinovsky for interesting and informative
discussions.

\bibliographystyle{amsplain}
\bibliography{trap}

\providecommand{\bysame}{\leavevmode\hbox to3em{\hrulefill}\thinspace}
\begin{thebibliography}{10}

\bibitem{AW:89}
A.~B. Aceves and S.~Wabnitz, \emph{Self induced transparency solitons in
  nonlinear refractive periodic media}, Phys. Lett. A \textbf{141} (1989),
  37--42.

\bibitem{A:95}
G.~P. Agrawal, \emph{Nonlinear fiber optics}, second ed., Academic Press, San
  Diego, 1995.

\bibitem{Agrawal}
\bysame, \emph{Fiber-optic communication systems}, Wiley--Interscience, 1997.

\bibitem{BPZ}
I.~V. Barashenkov, D.~E. Pelinovsky, and E.~V. Zemlyanaya, \emph{Vibrations and
  oscillatory instabilities of gap solitons}, Phys. Rev. Lett. \textbf{80}
  (1998), 5117--5120.

\bibitem{BD:98}
N.~G.~R. Broderick and C.~M. de~{S}terke, \emph{Approximate method for gap
  soliton propagation in nonuniform {B}ragg gratings}, Phys. Rev. E \textbf{58}
  (1998), 7941--7950.

\bibitem{BRI}
N.~G.~R. Broderick, D.~J. Richardson, and M.~Ibsen, \emph{Nonlinear switching
  in a 20-cm-long fiber {B}ragg grating}, Opt. Lett. \textbf{25} (2000),
  536--538.

\bibitem{CJ:89}
D.~N. Christodoulides and R.~I. Joseph, \emph{Slow {B}ragg solitons in
  nonlinear periodic structures}, Phys. Rev. Lett. \textbf{62} (1989),
  1746--1749.

\bibitem{DS:94}
C.~M. {d}e Sterke and J.~E. Sipe, \emph{Gap solitons}, Progress in Optics
  \textbf{33} (1994), 203--260.

\bibitem{E:97}
B.~J. Eggleton, C.~M. {d}e {S}terke, and R.~E. Slusher, \emph{Nonlinear pulse
  propagation in {B}ragg gratings}, J. Opt. Soc. Am. B. \textbf{14} (1997),
  no.~11, 2980--2992.

\bibitem{FKV:92a}
Z.~Fei, Y.~S. Kivshar, and L.~V\'{a}zquez, \emph{Resonant kink-impurity
  interactions in the $\phi^4$ model}, Phys. Rev. A \textbf{46} (1992),
  5214--5220.

\bibitem{FKV:92}
\bysame, \emph{Resonant kink-impurity interactions in the sine-{G}ordon model},
  Phys. Rev. A \textbf{45} (1992), 6019--6030.

\bibitem{FPM:94}
K.~Forinash, M.~Peyrard, and B.~Malomed, \emph{Interaction of discrete
  breathers with impurity modes}, Phys. Rev. E \textbf{49} (1994), 3400--3411.

\bibitem{GHW:01a}
R.~H. Goodman, P.~J. Holmes, and M.~I. Weinstein, \emph{Interaction of {NLS}
  solitons with defects: {P}hase space transport in a finite-dimensional
  model},  (preprint, 2001).

\bibitem{GHW:01}
\bysame, \emph{Interaction of sine-{G}ordon kinks with defects: {P}hase space
  transport in a two-mode model},  (preprint, 2001).

\bibitem{GWH:01}
R.~H. Goodman, M.~I. Weinstein, and P.~J. Holmes, \emph{Nonlinear propagation
  of light in one dimensional periodic structures}, J. of Nonlinear Sci.
  \textbf{11} (2001), 123--168.

\bibitem{HT}
A.~Hasegawa and F.~Tappert, \emph{Transmission of stationary nonlinear optical
  pulses in dispersive dielectric figers, {I}: {A}nomalous dispersion}, Appl.
  Phys. Lett. \textbf{23} (1971), 142--144.

\bibitem{M:85}
B.~A. Malomed, \emph{Inelastic interactions of solitons in nearly integrable
  systems 2}, Physica D \textbf{15} (1985), 385--401.

\bibitem{M_etal}
P.~Millar, R.~M. {D}e~{L}a {R}ue, T.~F. Krauss, J.~S. Aitchison, N.~G.~R.
  Broderick, and D.~J. Richardson, \emph{Nonlinear propagation effects in an
  {A}l{G}a{A}s {B}ragg grating filter}, Opt. Lett. \textbf{24} (1999),
  685--687.

\bibitem{MSG}
L.~F. Mollenauer, R.~H. Stolen, and J.~P. Gordon, \emph{Experimental
  observation of picosecond pulse narrowing and solitons in optical fibers},
  Phys. Rev. Lett \textbf{45} (1980), 1095--1098.

\bibitem{RW:88}
H.~Rose and M.~I. Weinstein, \emph{On the bound states of the nonlinear
  {S}chr\"odinger equation with a linear potential}, Physica D \textbf{30}
  (1988), 207--218.

\bibitem{SS:85}
G.-H. Song and S.-Y. Shin, \emph{Design of corrugated waveguide filters by the
  {G}el'fand-{L}evitan-{M}archenko inverse scattering method}, J. Opt. Soc. Am.
  A \textbf{2} (1985), 1905--1915.

\bibitem{T:62}
E.~C. Titchmarsh, \emph{Eigenfunction expansions associated with second-order
  differential equations. {P}art {I}}, Clarendon Press, Oxford, 1962.

\bibitem{W:99}
M.~I. Weinstein, \emph{Notes on wave propagation in 1-d periodic media with
  defects}, Tech. report, Bell Labs, 1999.

\bibitem{ZS1}
V.~E. Zakharov and A.~B. Shabat, \emph{Exact theory of two-dimensional
  self-focusing and one dimensional self-modulation of waves in nonlinear
  media}, Sov. Phys. JETP \textbf{34} (1972), 62--69.

\bibitem{ZS2}
\bysame, \emph{Interaction between solitons in a stable medium}, Sov. Phys.
  JETP \textbf{37} (1973), 823--828.

\end{thebibliography}

\end{document}